\documentclass[fleqn,usenatbib]{mnras}

\usepackage[T1]{fontenc}

\DeclareRobustCommand{\VAN}[3]{#2}
\let\VANthebibliography\thebibliography
\def\thebibliography{\DeclareRobustCommand{\VAN}[3]{##3}\VANthebibliography}
\defcitealias{NS10}{NS}
\defcitealias{Buder2020}{GALAH DR3}
\defcitealias{SDSSDR16}{APOGEE DR16}
\defcitealias{Fink2014}{F14}
\defcitealias{LN2018}{LN18}
\defcitealias{Seitenzahl2013}{S13}
\defcitealias{Shen2018}{S18}
\defcitealias{Bravo2019}{B19}
\defcitealias{Kobayashi2020}{K20}
\defcitealias{LeungNomoto2020}{LN20}

\usepackage{graphicx}	
\usepackage{amsmath}	
\usepackage{amssymb}	
\usepackage{paralist}
\usepackage{xspace}
\usepackage{pdflscape}
\usepackage{newtxtext,newtxmath}

\newcommand{\percent}{\mathrm{per\,cent}}
\newcommand{\tsfh}{\tau_\mathrm{sfh}}
\newcommand{\tia}{\tau_\mathrm{Ia}}
\newcommand{\tdep}{\tau_\mathrm{dep}}
\newcommand{\tstar}{\tau_\star}
\newcommand{\tD}{t_\mathrm{D}}
\newcommand{\e}{\mathrm{e}}
\newcommand{\dex}{\,\mathrm{dex}}
\newcommand{\gs}{Gaia \textit{Sausage}\xspace}
\newcommand{\xfe}[1]{[#1/Fe]\xspace}
\newcommand{\xfemath}[1]{[\mathrm{#1}/\mathrm{Fe}]}
\newcommand{\feh}{[Fe/H]\xspace}
\newcommand{\fehmath}{[\mathrm{Fe}/\mathrm{H}]}
\newcommand{\xh}[1]{[#1/H]\xspace}
\newcommand{\afe}{[$\alpha$/Fe]\xspace}
\newcommand{\norm}[2]{\sim\mathcal{N}(#1,#2)}
\newcommand{\teff}{T_\mathrm{eff}}
\newcommand{\logg}{\log g}
\newcommand{\ia}{SNe Ia\xspace}
\newcommand{\sCh}{sub-$M_\mathrm{Ch}$\xspace}
\newcommand{\Ch}{$M_\mathrm{Ch}$\xspace}

\title[Sub-Chandrasekhar Type Ia in the Gaia Sausage]{Evidence for sub-Chandrasekhar Type Ia supernovae from the last major merger}

\author[J.~L.~Sanders, V.~Belokurov and K.~T.~F.~Man]{
Jason L. Sanders$^{1}$\thanks{E-mail: jason.sanders@ucl.ac.uk (JLS)},
Vasily Belokurov$^{2}$
and 
Kai T.~F. Man$^{2}$
\\
$^{1}$Department of Physics and Astronomy, University College London, London WC1E 6BT, UK\\
$^{2}$Institute of Astronomy, University of Cambridge, Madingley Rise, Cambridge, CB3 0HA, UK
}

\date{Accepted XXX. Received YYY; in original form ZZZ}

\pubyear{2021}

\begin{document}
\label{firstpage}
\pagerange{\pageref{firstpage}--\pageref{lastpage}}
\maketitle

\begin{abstract}
We investigate the contribution of sub-Chandrasekhar mass Type Ia supernovae to the chemical enrichment of the Gaia \emph{Sausage} galaxy, the progenitor of a significant merger event in the early life of the Milky Way. Using a combination of data from Nissen \& Schuster (2010), the 3rd GALAH data release (with 1D NLTE abundance corrections) and APOGEE data release 16, we fit analytic chemical evolution models to a 9-dimensional chemical abundance space (Fe, Mg, Si, Ca, Cr, Mn, Ni, Cu, Zn) in particular focusing on the iron-peak elements, Mn and Ni. We find that low [Mn/Fe] $\sim-0.15\,\mathrm{dex}$ and low [Ni/Fe] $\sim-0.3\,\mathrm{dex}$ Type Ia yields are required to explain the observed trends beyond the [$\alpha$/Fe] knee of the Gaia \emph{Sausage} (approximately at [Fe/H] $=-1.4\,\mathrm{dex}$). Comparison to theoretical yield calculations indicates a significant contribution from sub-Chandrasekhar mass Type Ia supernovae in this system (from $\sim60\,\percent$ to $100\,\percent$ depending on the theoretical model with an additional $\pm10\,\percent$ systematic from NLTE corrections). We compare to results from other Local Group environments including dwarf spheroidal galaxies, the Magellanic Clouds and the Milky Way's bulge, finding the Type Ia [Mn/Fe] yield must be metallicity-dependent.
Our results suggest that sub-Chandrasekhar mass channels are a significant, perhaps even dominant, contribution to Type Ia supernovae in metal-poor systems, whilst more metal-rich systems could be explained by metallicity-dependent sub-Chandrasekhar mass yields, possibly with additional progenitor mass variation related to star formation history, or an increased contribution from Chandrasekhar mass channels at higher metallicity.
\end{abstract}

\begin{keywords}
Galaxy: abundances, evolution -- supernovae: general -- nucleosynthesis
\end{keywords}


\section{Introduction}
The progenitors of Type Ia supernovae (\ia) are not well understood. Despite this, there has been great success in their use as standard candles for measuring the accelerating expansion of the Universe \citep{Riess1998,Perlmutter1999}. This has been possible due to empirical relations between luminosity and light-curve decay rate allowing for accurate calibration of their standard candle nature \citep[e.g.][]{Pskovskii1977,Phillips1993}. However, the cause of such a relation is unclear. There is a growing zoo of different sub-classes of observed \ia, indicating a range of formation channels. The commonality between most observed \ia is they likely arise from the explosion of a carbon-oxygen (CO) white dwarf (WD) in a binary system \citep{Whelan1973,IbenTutukov1984,Webbink1984}. Beyond this, there are many suggested scenarios for the nature of the progenitor systems and the explosion mechanisms \citep[see the reviews of][]{HillebrandtReview, MaozReview2014,SeitenzahlTownsley2017,Ruiter2020}.

The two leading theories for the progenitors of normal \ia are the single- and double-degenerate scenarios. Single-degenerate progenitor systems are mostly anticipated to produce explosions near the Chandrasekhar mass (\Ch), whilst double-degenerate systems can produce explosions well below the Chandrasekhar mass (\sCh). In the single-degenerate scenario, a CO WD typically accretes from a non-degenerate H-rich companion (non-degenerate He donors are also possible and at low accretion rates could resemble the double-degenerate scenario described below) until the WD approaches the Chandrasekhar mass \citep{Whelan1973}. At first, the WD enters a turbulent low-level carbon-burning phase lasting a few hundred to a thousand years, which eventually ignites the dynamical burning of carbon. Early models of this scenario \citep{Arnett1969} considered a super-sonic detonation which failed to produce the intermediate-mass elements (e.g. Si) observed in \ia remnants due to the high central density of the WD \citep{Arnett1971}. To remedy this, typically a deflagration flame is initialized which propagates sub-sonically and causes the WD to expand, producing regions of low density \citep{Nomoto1984}. Eventually, the deflagration flame transitions to a detonation wave leading to the production of intermediate mass elements in the new lower density regions, and crucially suppressing the production of neutron-rich species (e.g. $^{58}$Ni) which are overproduced in pure deflagration models
\citep{Khokhlov1991}.
This scenario is dubbed the deflagration-to-detonation transition (DDT, sometimes called delayed detonation). 

It has further been hypothesised that in some systems the transition to detonation fails to occur so the WD undergoes a pure deflagration \citep[e.g.][]{Fink2014} leaving behind a remnant `zombie' core. These systems could explain the observed class of under-luminous Type Iax supernovae \citep{Foley2013}. Despite the success the single-degenerate DDT models have in explaining many properties of the observed light-curve and remnant \citep[e.g.][]{Hoeflich1996,Badenes2005}, a few issues remain:
\begin{inparaenum}
\item the stable accretion rate of hydrogen onto the WD surface must be within a relatively narrow range such that it can successfully reach the Chandrasekhar mass \citep[if the rate is too high a red giant envelope can form leading to common envelope evolution unless a strong wind is driven; if too low hydrogen flashes and nova explosions eject the mass,][]{Nomoto2018}, 
\item there is an observed lack of near-Chandrasekhar mass WDs from which these systems could originate \citep[e.g.][]{Kepler2007},
\item the predicted rates from this channel fall short of explaining the observed \ia rates (e.g. \citealt{Ruiter2009} and see figure 8 of \citealt{MaozReview2014} for a comparison),
\item low X-ray flux from nearby elliptical galaxies are consistent with only $5\,\percent$ of all \ia being single-degenerate \citep{Gilfanov2010},
\item the late-time spectrum of the Type Ia supernova SN2012cg lacks H$\alpha$ emission as expected from the envelope of a non-degenerate companion \citep{Shappee2018},
\item and there are still some inconsistencies in matching the colours, velocities and light curves in all bands \citep{Sim2013}.
\end{inparaenum}

The double-degenerate scenario instead considers two white dwarfs in a binary system (either both CO WDs, or a CO WD and a lower-mass He WD) slowly merging due to the emission of gravitational waves  \citep{IbenTutukov1984}. Although early work presumed the disrupting WD would form an accretion disc around the primary \citep{Benz1990} leading to steady growth up to \Ch, it is now deemed more probable that one WD initialises a detonation in the other well before the Chandrasekhar mass is reached. Lower central densities in the \sCh WDs lead naturally to production of intermediate-mass elements without invoking a deflagration phase. If He is accreted through Roche-lobe overflow from a companion WD at a low enough rate to avoid helium flashes, a sufficiently massive He envelope can build up to initiate a detonation at the base of the He layer, triggering a detonation in the carbon core \citep{WoosleyWeaver1994}. This scenario is called double-detonation (double-detonations may also occur in single-degenerate systems with a non-degenerate He-star although these are expected to have short delay times so cannot explain the bulk of \ia with $>1\,\mathrm{Gyr}$ delays, and the initial detonation cannot be dynamically triggered, see below). The mass of the He shell required to initialise a detonation ($\sim0.2M_\odot$) produces high-velocity iron-peak elements from the burning of the He shell not seen in observations \citep{Hoeflich1996} and an overabundance of lighter iron-peak elements \citep[e.g. V and Cr,][]{Kobayashi2020}. However, carbon detonation can occur with lower mass He shells \citep[e.g.][]{Fink2010} with the He detonation possibly being triggered dynamically through unstable mass transfer from a violent merger \citep{Guillochon2010,Pakmor2012,Shen2018}. This scenario produces superior matches to observed multi-band light-curves \citep{Sim2010,Kromer2010,Townsley2019} and also predicts the ejection of the WD companion at high velocities \citep[see][for potential candidates from Gaia]{Shen2018B}. 

One route for constraining the contributions of these different \ia formation scenarios is through galaxy chemical evolution modelling. \ia produce $\sim0.6M_\odot$ of iron (formed from the radioactive decay of $^{56}$Ni which causes \ia to glow) and low (although as discussed not insignificant) quantities of intermediate-mass $\alpha$ elements. The contribution of \ia can then be inspected from the down-turning of a stellar population in the \afe-\feh diagram \citep{Tinsley1979}. Although theoretical models for the different progenitor scenarios agree in the iron production per event, because of the variety of densities in the exploding media there is a range of different chemical abundances produced, particularly for the iron-peak elements. Most notably Mn \citep[e.g.][]{Seitenzahl2013B} and Ni \citep[e.g.][]{Kirby2019} are very sensitive to the central density of the exploding WD \citep[see also][]{McWilliam2018}. Mn is mostly produced through the decay of $^{55}$Co which is made in large quantities in normal freeze-out from nuclear statistical equilibrium achieved in \Ch WDs, or in lower quantities in incomplete silicon burning at the lower densities typical in \sCh WDs, and is destroyed during alpha-rich freeze-out which occurs at lower densities \citep[see][]{Lach2020}. Stable Ni is produced as $^{58}$Ni which arises from the neutron-rich environments produced by electron-capture in the high density environments of \Ch WDs \citep{SeitenzahlTownsley2017}. For these reasons, typically \sCh explosions lead to sub-solar \xfe{Mn} and \xfe{Ni}, whilst \Ch events produce super-solar yields. Other elements (e.g. Cr, V) are also sensitive indicators of the explosion properties \citep{Palla2021}. Studies of the \xfe{Mn} distribution in the Milky Way have thus concluded a significant fraction ($\gtrsim75\,\percent$) of Chandrasekhar mass supernovae is required to explain the observations \citep{Seitenzahl2013B,Kobayashi2020} although comparative studies of different host environments \citep[the bulge, thick disc and dSphs, see][]{Cescutti2008,North2012} have appealed to significantly metallicity-dependent \ia yields. A significant fraction of \sCh systems is in line with the ejecta mass measurements from \cite{Scalzo2014} who suggest $25-50\,\percent$ of systems are inconsistent with \Ch explosions, and \cite{Flors2020} who suggest $85\,\percent$ of spectroscopic Ni observations of \ia are consistent with \sCh models. Recently, \cite{Kirby2019} and \cite{delosReyes2020} have used Ni and Mn abundance measurements respectively for dwarf spheroidal galaxies (most notably Sculptor) to argue sub-solar \xfe{Mn} and \xfe{Ni} \ia yields are needed, potentially indicating a high fraction of sub-Chandrasekhar mass events in these low-metallicity or early star-forming systems \citep{Kobayashi2020}. Furthermore, \cite{McWilliam2018} have used the low \xfe{Mn} and \xfe{Ni} abundances of a single metal-rich star in Ursa Minor to argue the star has been significantly enriched by a single sub-Chandrasekhar mass SNe Ia event. To build a consistent model of \ia progenitors, we require a range of systems with different star formation histories and different metallicities which we can simultaneously fit. 

With the arrival of data from the \emph{Gaia} satellite \citep{Gaia}, along with accompanying spectroscopic surveys, significant evidence has emerged that the Milky Way experienced a $>10^{10}M_\odot$ (total mass) merger approximately $8-10\,\mathrm{Gyr}$ ago. The progenitor system is known as the \gs or Gaia-Enceladus. \cite{Evans2020} outlines the history of this idea. \cite{NS10} discovered two populations of local halo dwarf stars with distinct kinematics which follow distinct chemical abundance tracks, and attributed the lower \afe sequence to a remnant dwarf galaxy. \citet{Deason2013} put forward the hypothesis that the prominent break in the Galactic stellar halo's density profile is the consequence of an ancient accretion of a substantial satellite galaxy. This idea was tested by \cite{Belokurov2018} who used early Gaia and SDSS data to show that the bulk of the local Milky Way's halo (corresponding to intermediate metallicities $\fehmath\sim-1.2\dex$) is contributed by stars on very radial orbits. Comparing the Gaia-SDSS observations to cosmological simulations of MW-like galaxy formation, \cite{Belokurov2018} concluded that such a dominant, metal-rich and highly-eccentric halo component is naturally produced in a collision between the MW and an LMC/SMC-mass galaxy around redshift $z\sim2$.  \cite{Helmi2018} and \citet{Mackereth2019} presented a coherent chemo-dynamical picture of the merger using APOGEE data in combination with Gaia DR2, demonstrating that the progenitor's chemical evolution history can be reconstructed robustly using a set of simple orbital selection criteria. Further evidence for the merger has come from the properties of the globular cluster population \citep[e.g.][]{Myeong2018}. The \gs merger presents an ideal opportunity to study chemical evolution in a low metallicity, ``high redshift'' galaxy using a sample of bright, nearby stars. Although the galaxy is no longer intact, we can still study the distributions of its constituent stars in chemical abundance space and make inferences on the evolution within the progenitor system \citep[e.g.][]{Vincenzo2019,Aguado2020,Matsuno2021}. The system nicely bridges the gap between the lower metallicity dSph systems studied by \citet{Kirby2019} and \citet{delosReyes2020} and the higher metallicity environments within the Milky Way populations, whilst having abundant high-quality data. Furthermore, it is perhaps the best nearby example of a reasonably high-mass galaxy with a truncated early burst of star formation so presents a unique opportunity to study chemical evolution in this setting. With the current influx of high-dimensional abundance data from large stellar spectroscopic surveys such as APOGEE \citep{SDSSDR16} and GALAH \citep{Buder2020}, it is now possible to perform the necessary detailed studies of the chemical properties of this structure.

In this paper, we use a high-dimensional chemical abundance space to constrain simple chemical evolution models of the \gs and place constraints on the properties of Type Ia supernovae in this system. We discuss the data used in Section~\ref{sec::data}. In Section~\ref{sec::method} we describe the models from \cite{Weinberg2017} used to fit the data. Focussing specifically on the measured Mn and Ni yields, we discuss our results compared to a range of theoretical yields from \Ch and \sCh Type Ia supernovae in Section~\ref{sec::discussion}, along with the impact of the assumption of local thermal equilibrium on our results. Finally, we compare the constraints on the Type Ia supernovae channel with results from other systems (the Milky Way and dwarf spheroidal galaxies) highlighting the necessary metallicity dependence of the yields, before we present our conclusions in Section~\ref{sec::conclusions}.

\section{Multi-dimensional chemical abundance data}\label{sec::data}
We employ three samples of data: the local metal-poor `halo' sample from \citet{NS10}, GALAH DR3 \citep{Buder2020} and APOGEE DR16 \citep{SDSSDR16}. We describe the properties of each of the datasets in turn as well as the cuts we have employed to produce samples of \gs member stars.

\subsection{Nissen \& Schuster (2010,11)}

First, we take the nearby (distance $<335\,\mathrm{pc}$) sample of low metallicity dwarf stars from \citet[][NS]{NS10} and \cite{NS11}. The stars were observed with either the UVES spectrograph on the VLT ($R\sim55000$) or the FIES spectrograph on the NOT ($R\sim40000$) and LTE abundance measurements were obtained from equivalent widths for Fe, Na, Mg, Si, Ca, Ti, Cr, Mn, Cu, Zn, Ni, Y and Ba. 
For these stars we further utilise 
the non-LTE Cu measurements from \cite{Yan2016}. We cross-match to the Gaia Early Data Release 3 \citep[EDR3,][]{Gaia,GaiaEDR3} to obtain proper motions and distances (utilizing a simple parallax inversion with a $0.017\,\mathrm{mas}$ offset), and find orbital properties (eccentricity $e$, maximum Galactic height $z_\mathrm{max}$) using \textsc{Galpy} \citep{galpy} using the \cite{McMillan2017} potential. We separate the \gs stars from the \emph{in-situ} population by cutting on eccentricity \citep[\gs stars have $e>0.7$,][although a stricter cut of $e>0.85$ does not change the subsequent results significantly]{Mackereth2019,Naidu2020} and in the (\feh, \xfe{Mg}) space (\gs stars have \xfe{Mg}$<-0.35\fehmath-0.07$). This results in $38$ \gs stars. The highest metallicity of the sample is $-0.75\,\dex$ making an additional maximum metallicity cut (as done for the GALAH and APOGEE samples) unnecessary. We adopt the statistical uncertainties $\Delta$[X/Fe] provided by \citetalias{NS10} and \cite{NS11} (estimated from repeat observations) of (Mg, Si, Ca, Cr, Mn, Ni, Cu, Zn)$=(0.03,0.03,0.02,0.02,0.025,0.01,0.035,0.035)\dex$ and $\Delta$[Fe/H]$=0.04\dex$.

\subsection{GALAH DR3}\label{sec::data_galah}

Secondly, we utilise the 3rd data release (DR3) of the GALAH survey \citep{Buder2020,Kos2017}. GALAH (Galactic Archaeology with HERMES) is a large stellar spectroscopic survey performed with the HERMES spectrograph \citep{HERMES} on the Anglo-Australian Telescope. GALAH targets primarily stars with $9<V<14$ although the recent data release has processed similar surveys performed with HERMES, some of which are fainter. The majority of the targetted stars are within a few kpc of the Sun. In DR3, 30 elemental abundances are provided along with uncertainties, quality flags and bitmasks indicating the lines used for each star. 11 of these elements have 1D non-local thermal equilibrium (LTE) corrections \citep{Amarsi2020}. Furthermore, \citetalias{Buder2020} provides several value-added catalogues including the computation of spectro-photometric distances incorporating Gaia DR2 data and orbital parameters \citep[computed in the Galactic potential of][]{McMillan2017}. We restrict to stars with $-2\dex<$ \feh$<-0.7\dex$ (with \texttt{flag\_sp} $=0$, \texttt{flag\_fe\_h} $=0$ and \texttt{flag\_alpha\_fe} $=0$) and select a sample of high-probability \gs members by requiring $\mathrm{med}(e)>0.85$, $\Delta e<0.05$, \xfe{Mg}$<-0.2$\feh and \xfe{Si}$<0.42\dex$. The cut in the \xfe{Mg} vs. \feh plane is different here compared to that adopted for \citetalias{NS10} and \citetalias{SDSSDR16} due to the differences in the \xfe{Mg} abundance scales discussed in Appendix~\ref{appendix::common_stars}. The stricter eccentricity cut compared to that employed for \citetalias{NS10} is to reduce contamination from other accreted substructures \citep{Myeong2019,Kim2021}. The \xfe{Si} cut removes obvious contaminants not observed in the other datasets. We remove stars within $3$ half-light radii (in projection) of known globular clusters \citep[][2010 version]{HarrisGC} and within known open clusters \citep{Dias2002}. Furthermore, we only use the abundances with uncertainties smaller than $0.1\dex$ and with no flags. This results in a sample of $1047$ stars. 

In Appendix~\ref{appendix::common_stars} we inspect the abundance differences with respect to APOGEE DR16. We find significant offsets and trends with metallicity for many abundances. Part of this could arise from the non-LTE corrections applied in GALAH. For Ni (which does not have a non-LTE correction in GALAH), two lines are used in GALAH DR3. We find a significant difference in the residuals for stars analysed using both lines compared to when only one is used. We therefore choose to apply a correction to the GALAH DR3 Ni abundances to put them on the APOGEE scale. 

\subsection{APOGEE DR16}\label{sec::data_apogee}

Thirdly, we use the APOGEE survey data as part of SDSS data release 16 \citep[DR16,][]{SDSSDR16}. APOGEE \citep[Apache Point Observatory Galactic Evolution Experiment,][]{Majewski2017} is a large infra-red spectroscopic survey which has taken observations of $\sim430,000$ stars in the $H$ band at resolution $R\sim22,500$. The initial survey (APOGEE-1) used the $2.5$m Sloan Foundation Telescope \citep{Gunn2006} situated at Apache Point Observatory, but from DR16 the survey was extended (APOGEE-2) to include additional data taken with the $2.5$m du Pont Telescope situated at Las Campanas Observatory. Both sub-surveys used similar spectrographs \cite{Wilson2019}. Abundances for $\sim20$ elements are provided as measured by the ASPCAP pipeline \citep{aspcap,Jonsson2018,Jonsson2020} along with uncertainties and quality flags. We only use data with \texttt{aspcapflag} $=0$ (and as with GALAH remove stars near known globular and open clusters). Unlike the GALAH survey, all abundances are derived under an LTE approximation, although non-LTE corrections are expected in the next data release \citep{Osorio2020}. \cite{LeungBovy2019} have provided a value-added catalogue with distances and orbital properties computed for all APOGEE stars via a convolutional neural network (astroNN). This catalogue assumes the MWPotential2014 from \cite{galpy} whilst for the other datasets we have assumed the \cite{McMillan2017} potential. For stars matching between APOGEE and GALAH with eccentricity uncertainties less than $0.1$, the two catalogues give a median eccentricity difference of $0.01$ with a scatter of $0.06$ so the difference between the Galactic potentials is not a concern.
We select stars with $\fehmath<-0.7\dex$, $\Delta\fehmath<0.05\dex$, $\Delta\xfemath{Mg}<0.05\dex$, $\xfemath{Mg}<-0.35\fehmath-0.07$, $e>0.85$, $\Delta e<0.05$ and only use non-flagged abundances with uncertainties $<0.05\dex$ ($0.1\dex$ for Cr and Cu). Abundances are not provided for low \xh{X} which can lead to biases at low metallicities. This is most apparent for \xh{Mn} which has a floor at $-2.25\,\dex$. This produces an upturn in the mean trend of \xfe{Mn} vs \feh below $\xfemath{Mn}\sim-1.5\dex$. Our strict uncertainty cut removes \xfe{Mn} measurements below this value leading to unbiased results. The effect is weaker for the other abundances but again our cuts avoid any biases it might introduce. Our final sample consists of $819$ stars. 

\subsection{Common stars}\label{sec:common}
In Appendix~\ref{appendix::common_stars} we have analysed the differences in abundances for stars common to the three presented samples. When comparing \citetalias{SDSSDR16} to \citetalias{NS10} we find only \xfe{Si} has a significant offset with \citetalias{SDSSDR16} $0.09\dex$ lower than \citetalias{NS10}. The only significant offset between \citetalias{Buder2020} and \citetalias{NS10} is in \xfe{Cr} which is $0.11\dex$ lower in \citetalias{Buder2020} although there are only a few stars in common. When comparing \citetalias{SDSSDR16} and \citetalias{Buder2020} the differences arising from the non-LTE corrections of Mg, Si and Ca from \cite{Amarsi2020} are apparent. Mn has non-LTE corrections in GALAH but there is no significant systematic trend with respect to the APOGEE Mn abundances detected. As discussed in Section~\ref{sec::data_galah}, the GALAH Ni abundances appear to be systematically biased at low metallicity compared to the APOGEE Ni abundances.

\section{Analytic chemical evolution modelling}\label{sec::method}
\cite{Weinberg2017} demonstrate how galactic chemical evolution can be solved analytically under a number of simplifying assumptions:
\begin{inparaenum}
\item the galaxy is represented by a single `zone' so no radial flows or radial migration are included,
\item the star formation efficiency (the star formation per unit gas mass $\equiv1/\tstar$) is a constant (instead of scaling non-linearly as per a Kennicutt-Schmidt law),
\item supernovae (or other enrichment sources) either return their products immediately (e.g. for Type II) or, for Type Ia, according to an exponential delay-time distribution (DTD), $\e^{-(t-\tD)/\tia}$, with timescale $\tia$ and minimum delay time $\tD$ (or a linear combination of such distributions),
\item stellar yields are independent of initial metallicity,
\item the rate of gas outflow is proportional to the star formation rate, via a constant $\eta$, and finally,
\item the star formation history, $\dot M_\star(t)$, is from a limited set of functions including exponential $\dot M_\star\propto \e^{-t/\tsfh}$ or linear-exponential $\dot M_\star\propto t\e^{-t/\tsfh}$ for constant $\tsfh$.
\end{inparaenum}
In these models, the gas reservoir is depleted on a timescale $\tdep=\tstar/(1-r+\eta)$ where $r\approx0.4$ is the fraction of unprocessed material returned by stars \citep{Weinberg2017}. The advantage of these models is the speed with which they can be evaluated as they avoid a costly forward integration. 

For an assumed set of (IMF-integrated) net yields $m^\mathrm{Y}_{j}$ of element $j$ (i.e. mass of element produced per mass of stellar generation formed) and assuming a linear-exponential star formation law, the solution over time for the enrichment due to Type II supernovae ($\mathrm{Y}=\mathrm{II}$) is given by equation (56) of \cite{Weinberg2017}, whilst the enrichment due to Type Ia supernovae ($\mathrm{Y}=\mathrm{Ia}$) is given by equation (58). By normalizing these equations relative to an assumed solar abundance \citep{Asplund2009}, we are able to predict the abundance tracks [X/H]$(t)$ given a set of model parameters, $p=(\tsfh, \tstar, \eta, \tia, \tD)$. 

The observed DTD for \ia follows a power-law of the form $\sim t^{-1}$ above some minimum delay time, $\tD$ \citep{MaozMannucci2012}. As described in \cite{Weinberg2017}, a linear combination of exponential DTDs, $\sum_{k=1}^{N_\mathrm{e}} w_k\e^{-(t-\tD)/\tau_\mathrm{Ia,k}}$, can be used to approximate a power-law distribution over the times of interest. For each choice of $t_D$ we have fitted a combination of $N_\mathrm{e}=3$ exponential DTDs to $1/t^{1.1}$ normalized between $t_D$ and a Hubble time. We have then approximated the relationship between $t_D$ and the fitted timescales, $\tau_\mathrm{Ia,k}$, and relative weights, $w_k$, using cubic polynomials in the logarithms of these quantities. In this way, given a choice of $t_D$ we can simply find the combination of exponential DTDs which reproduces the $1/t^{1.1}$ DTD. Our parameter set $p$ is then reduced to $p=(\tsfh, \tstar, \eta, \tD)$. 

The key parameters for controlling the shape of the \xfe{X} vs. \feh tracks are $\tsfh$, $\tstar$ and $\tdep$. Increasing $\tsfh$ (slower enrichment), increasing $\tstar$ (less enrichment per unit gas mass) or decreasing $\tdep$ (increasing $\eta$, more outflow) all cause the tracks in $\xfemath{\alpha}$ vs. \feh to turn over at lower \feh. In the instantaneous recycling approximation, a metallicity-dependent Type II yield is equivalent to changing the depletion timescale, $\tdep$, (higher yields at higher metallicity increase the effective depletion timescale). Metallicity-dependent Type Ia yields are not possible within the analytic framework but would have a similar effect. Over the range of considered metallicities, theoretical yields depend only very weakly on metallicity (see later discussion). 

We express this model in \textsc{Pystan}, a \textsc{Python} interface to the probabilistic programming language \textsc{Stan} \citep{stan} for performing Bayesian inference. For each star $i$, we employ a birth time hyperparameter $t_i<t_\mathrm{max}$ from which we compute \{[X/Fe]$(t_i)$\} and [Fe/H]$(t_i)$ to compare with the data. Assuming selection effects are minimal, the $t_i$ follow the star formation rate, $t_i\sim\dot M_\star$, up to the maximum time $t_\mathrm{max}$ \citep[when the \gs merged with the Milky Way and star formation ceased,][, or more precisely in our analysis, the time when the interstellar gas metallicity reaches the maximum metallicity, $\fehmath\sim-0.7\,\dex$, of our defined samples]{Belokurov2020,Bonaca2020}. We marginalize over the model parameters $p=(\tsfh, \tstar, \eta, \tD, t_\mathrm{max})$ adopting normal distributions\footnote{Adopting the notation $\mathcal{N}(\mu,\sigma)$ for a normal distribution with mean $\mu$ and variance $\sigma^2$.} $\tD\norm{0.15}{0.05}$ (with a minimum at $0.04\,\mathrm{Gyr}$ corresponding to the lifetime of a $8M_\odot$ star assuming all more massive stars form Type II supernovae), $\tstar\norm{1}{100}$, $\eta\norm{10}{50}$, $t_\mathrm{max}\norm{4}{1}$ and $\tsfh\norm{2}{1}$ 
(where all timescales are in Gyr). We place broad priors on  $\xfemath{X}_\mathrm{Y}\norm{0}{2}$ except for $\xfemath{Mg}_\mathrm{Ia}$ which should be small \citep[e.g.][]{KobayashiChemicalEvolution} so we adopt a prior $\xfemath{Mg}_\mathrm{Ia}\norm{-2}{0.4}$ with a hard upper limit of $\xfemath{Mg}_\mathrm{Ia}=-0.5$ (which approximately corresponds to the range of theoretical Type Ia models considered later). Early models (particularly those fitted to the GALAH data) produced $\xfemath{Mg}_\mathrm{Ia}\approx0$ which is inconsistent with basic understanding of chemical evolution and essentially all \ia models. For the mass of iron produced from Type II supernovae per total mass of stars formed, we use $\ln m^\mathrm{II}_\mathrm{Fe}\norm{\ln 0.00075}{0.45}$, which is consistent with a range of net IMF-integrated yields (evaluated at $10^{-1.5}Z_\odot$) from the models of \cite{ChieffiLimongi2004}, \cite{Kobayashi2006},
\cite{LimongiChieffi2018} and NuGrid \citep{Nugrid1,Nugrid2} using IMFs from \cite{Salpeter1955},
\cite{Kroupa2001},
\cite{KroupaToutGilmore} and \cite{Chabrier2003} for stars between $0.1$ and $100M_\odot$. Similarly, for the range of Type Ia theoretical models considered later we find the mass of iron produced per event is $(0.77\pm0.15)M_\odot$. The total number of Type Ia supernovae over a Hubble time per stellar mass of star formation is approximately $(2.2\pm0.3)\times10^{-3}M_\odot^{-1}$ \citep{MaozMannucci2012}. Combining these numbers gives $\ln m^\mathrm{Ia}_\mathrm{Fe}\norm{\ln 0.0017}{0.24}$.
In addition to the reported abundance uncertainties, we also allow for additional scatter in all \xfe{X} using a two-component mixture model with mixture weight $f_\mathrm{mix}\norm{1}{0.05}$ (where $f_\mathrm{mix}$ is the weight of narrower component restricted to $0<f_\mathrm{mix}<1$) and priors $\ln\boldsymbol{\sigma}_1\norm{-3}{1}$ and $\ln\boldsymbol{\sigma}_2\norm{-1}{1}$ for the additive additional scatters. We also use the mixture for \feh but set $\sigma_1=0$.

\begin{figure*}
    \centering
    \includegraphics[width=\textwidth]{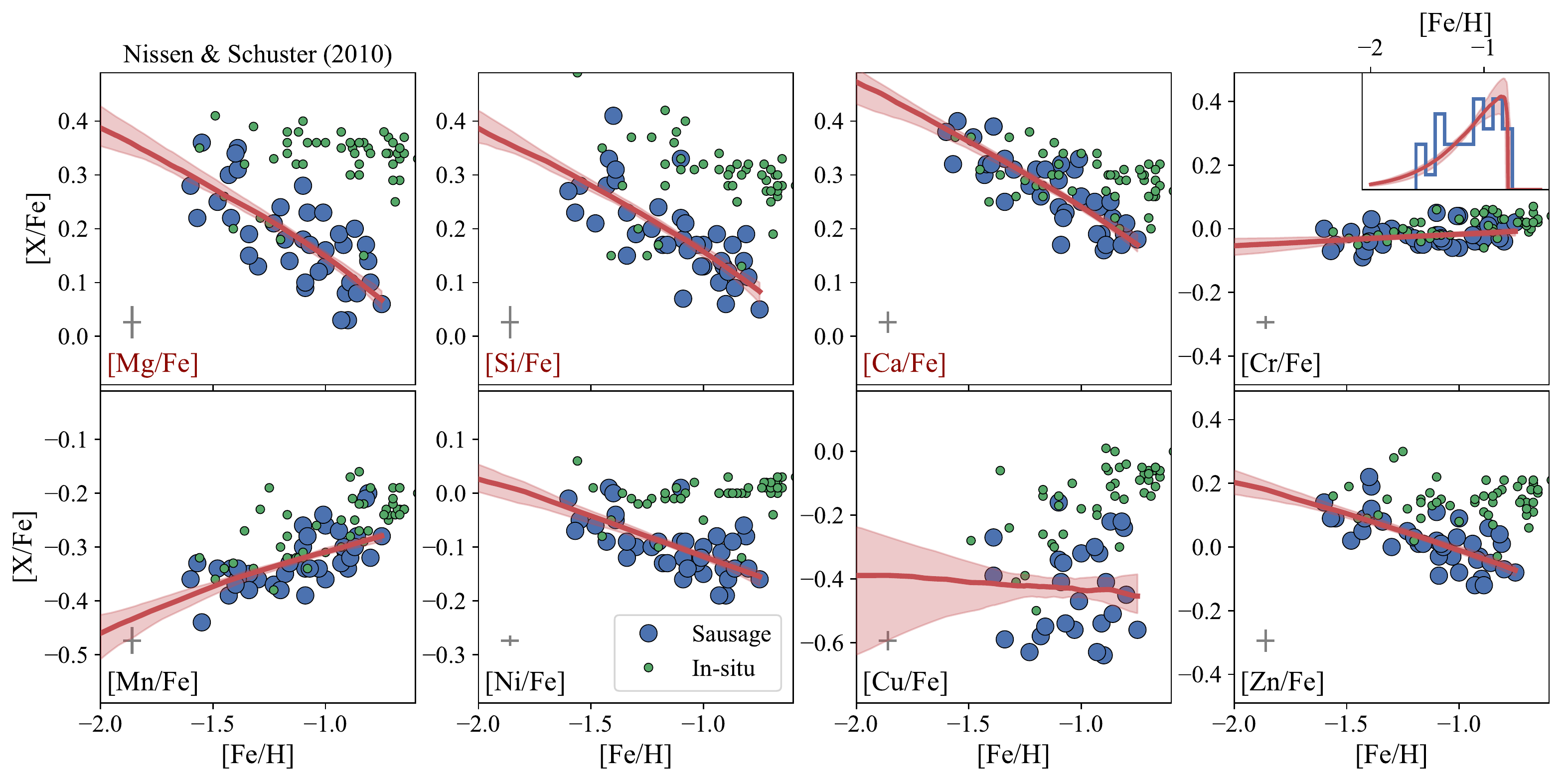}
    \caption{Analytic chemical evolution model fits to the \protect\cite{NS10} dataset. Large blue points are stars kinematically and chemically selected to belong to the \gs, whilst small green points are presumed in-situ Milky Way members. The red line and band show the fitted analytic chemical evolution models of \protect\cite{Weinberg2017} with the uncertainty. The inset errorbar shows the typical uncertainty in the abundances. The y-axis range spans $0.6\,\dex$ for all elements bar Cr, Cu and Zn where a range of $1\,\dex$ is used. The top right inset shows the 1D metallicity distribution and the model density. Note how even whilst \xfe{Mg} is declining with \feh, \xfe{Mn} stays relatively flat indicating the low \xfe{Mn} contribution from the Type Ia supernova channel.}
    \label{fig:ns_results}
\end{figure*}
\begin{figure*}
    \centering
    \includegraphics[width=\textwidth]{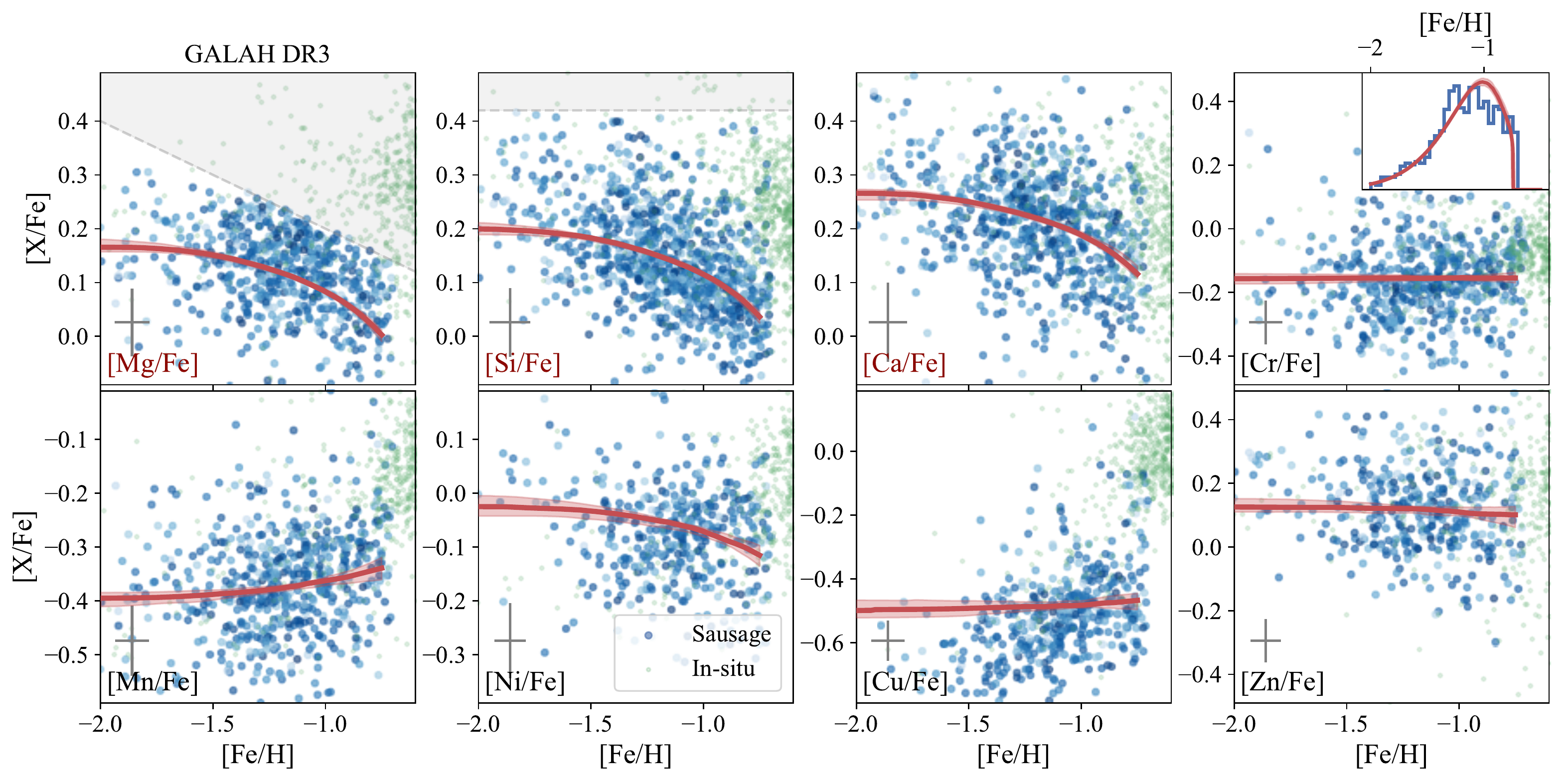}
    \caption{Analytic chemical evolution model fits to the \protect\citetalias{Buder2020} dataset ($\alpha$ element labels coloured red). Blue points are stars kinematically and chemically selected to belong to the \gs with darker colours corresponding to stars on more eccentric orbits, whilst small green points are presumed in-situ Milky Way members. The red line and band show the fitted analytic chemical evolution models of \protect\cite{Weinberg2017} with the uncertainty. The inset errorbar shows the typical uncertainty in the abundances.  The top right inset shows the 1D metallicity distribution and the model density. 1D non-LTE corrections are applied to Mg, Si, Ca and Mn \protect\citep{Amarsi2020}.}
    \label{fig:galah}
\end{figure*}
\begin{figure*}
    \centering
    \includegraphics[width=\textwidth]{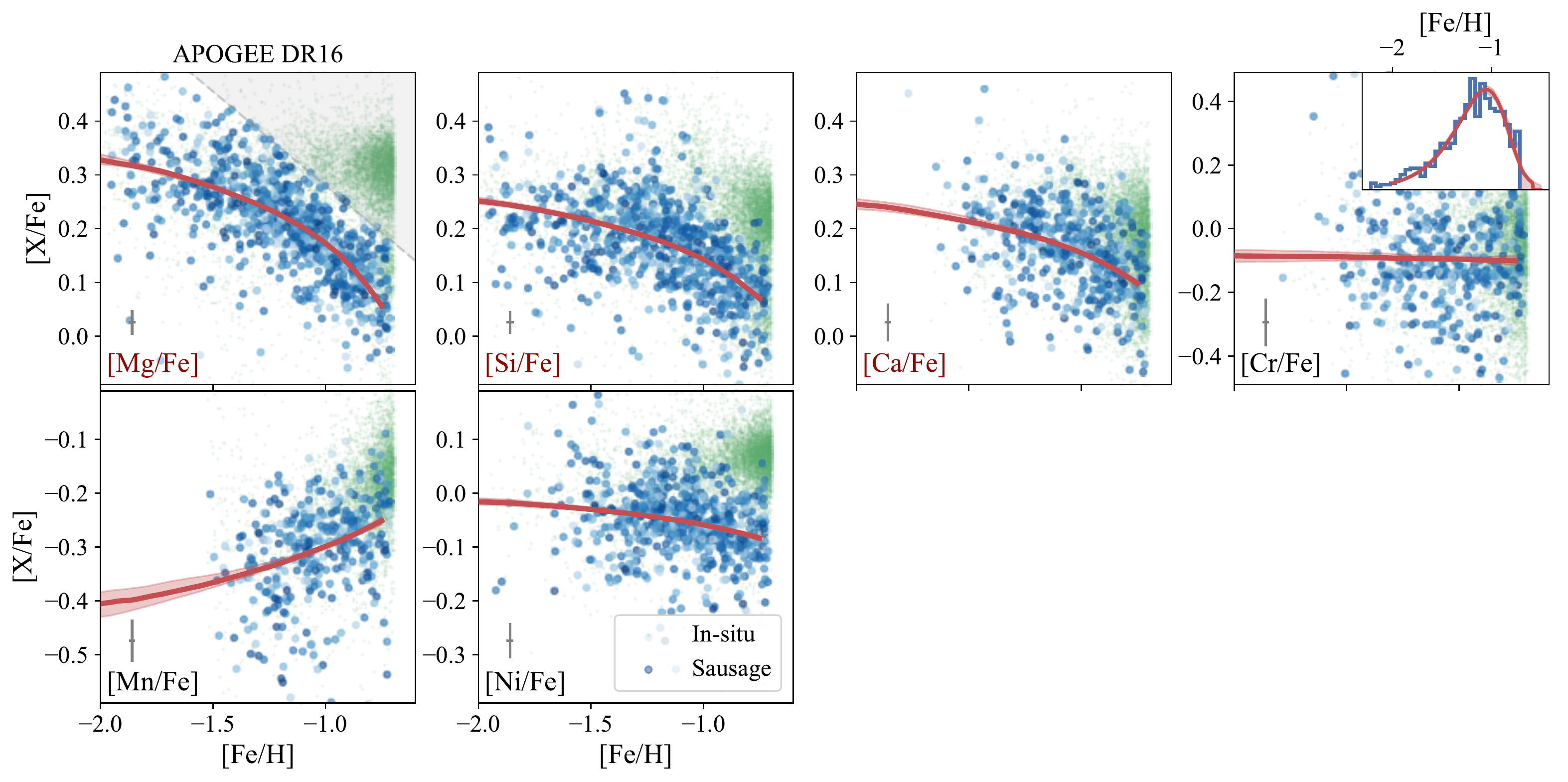}
    \caption{Analytic chemical evolution model fits to the \protect\citetalias{SDSSDR16} dataset ($\alpha$ element labels coloured red). See caption of Fig.~\protect\ref{fig:galah} for other details.}
    \label{fig:apogee}
\end{figure*}
\begin{table}
    \caption{Constraints and priors for parameters of the analytic chemical evolution models. All timescales are in Gyr.}
    \centering
    \begin{tabular}{lllll}
Param.&Prior&NS10&GALAH DR3&APOGEE DR16\\\hline\\
$\tau_\mathrm{sfh}$&$2\pm1$&$2.4^{+0.7}_{-0.6}$&$1.0^{+0.2}_{-0.2}$&$0.9^{+0.3}_{-0.2}$\\
$t_\mathrm{max}$&$4\pm1$&$4.4^{+0.9}_{-0.9}$&$3.9^{+1.0}_{-1.0}$&$5.1^{+0.8}_{-1.0}$\\
$\tau_\star$&$1\pm100$&$17.6^{+7.0}_{-5.5}$&$21.8^{+8.9}_{-6.8}$&$15.4^{+5.5}_{-4.0}$\\
$\eta$&$10\pm50$&$9.8^{+4.9}_{-3.9}$&$31.7^{+9.1}_{-6.9}$&$28.0^{+7.3}_{-5.8}$\\
$\tau_\mathrm{dep}$&$-$&$1.7^{+1.1}_{-0.6}$&$0.7^{+0.2}_{-0.2}$&$0.5^{+0.2}_{-0.1}$\\
$t_\mathrm{D}$&$0.15\pm0.05$&$0.13^{+0.05}_{-0.05}$&$0.18^{+0.05}_{-0.05}$&$0.08^{+0.09}_{-0.03}$\\
$f_\mathrm{mix}$&$1.00\pm0.05$&$1.00^{+0.00}_{-0.01}$&$0.86^{+0.02}_{-0.02}$&$0.90^{+0.01}_{-0.02}$\\
    \end{tabular}
    \label{tab:results_parameters}
\end{table}

\begin{table}
    \caption{Abundance constraints from analytic chemical evolution model fits. We report the (net IMF-integrated) mass per solar mass of star formation for Fe and the abundance relative to Fe for other elements. Type Ia yields are given in the top half of the table and Type II in the bottom half.}
    \centering
    \begin{tabular}{lllll}
Type&Element&NS10&GALAH DR3&APOGEE DR16\\\hline
Ia&$\log_{10}m_\mathrm{Fe}$&$-2.73^{+0.10}_{-0.10}$&$-2.89^{+0.09}_{-0.09}$&$-2.85^{+0.09}_{-0.09}$\\
Ia&$[\mathrm{Mg}/\mathrm{Fe}]$&$-2.00^{+0.35}_{-0.34}$&$-1.92^{+0.39}_{-0.37}$&$-1.96^{+0.36}_{-0.36}$\\
Ia&$[\mathrm{Si}/\mathrm{Fe}]$&$-0.87^{+0.34}_{-0.54}$&$-1.44^{+0.38}_{-0.51}$&$-0.31^{+0.05}_{-0.06}$\\
Ia&$[\mathrm{Ca}/\mathrm{Fe}]$&$-0.72^{+0.31}_{-0.52}$&$-0.78^{+0.33}_{-0.55}$&$-0.17^{+0.05}_{-0.05}$\\
Ia&$[\mathrm{Cr}/\mathrm{Fe}]$&$+0.03^{+0.03}_{-0.03}$&$-0.16^{+0.07}_{-0.09}$&$-0.12^{+0.05}_{-0.06}$\\
Ia&$[\mathrm{Mn}/\mathrm{Fe}]$&$-0.16^{+0.03}_{-0.03}$&$-0.24^{+0.05}_{-0.06}$&$-0.12^{+0.02}_{-0.02}$\\
Ia&$[\mathrm{Ni}/\mathrm{Fe}]$&$-0.41^{+0.08}_{-0.12}$&$-0.37^{+0.16}_{-0.26}$&$-0.17^{+0.02}_{-0.02}$\\
Ia&$[\mathrm{Cu}/\mathrm{Fe}]$&$-0.46^{+0.19}_{-0.35}$&$-0.42^{+0.10}_{-0.14}$&$-$\\
Ia&$[\mathrm{Zn}/\mathrm{Fe}]$&$-0.91^{+0.31}_{-0.52}$&$-0.01^{+0.15}_{-0.24}$&$-$\\
\hline
II&$\log_{10}m_\mathrm{Fe}$&$-3.08^{+0.16}_{-0.16}$&$-2.53^{+0.10}_{-0.10}$&$-2.71^{+0.10}_{-0.09}$\\
II&$[\mathrm{Mg}/\mathrm{Fe}]$&$+0.48^{+0.11}_{-0.09}$&$+0.17^{+0.01}_{-0.01}$&$+0.34^{+0.02}_{-0.02}$\\
II&$[\mathrm{Si}/\mathrm{Fe}]$&$+0.46^{+0.10}_{-0.09}$&$+0.20^{+0.01}_{-0.01}$&$+0.26^{+0.01}_{-0.01}$\\
II&$[\mathrm{Ca}/\mathrm{Fe}]$&$+0.54^{+0.10}_{-0.08}$&$+0.27^{+0.01}_{-0.01}$&$+0.25^{+0.02}_{-0.01}$\\
II&$[\mathrm{Cr}/\mathrm{Fe}]$&$-0.07^{+0.04}_{-0.06}$&$-0.15^{+0.02}_{-0.02}$&$-0.08^{+0.03}_{-0.03}$\\
II&$[\mathrm{Mn}/\mathrm{Fe}]$&$-0.56^{+0.11}_{-0.27}$&$-0.40^{+0.01}_{-0.01}$&$-0.42^{+0.03}_{-0.03}$\\
II&$[\mathrm{Ni}/\mathrm{Fe}]$&$+0.08^{+0.08}_{-0.06}$&$-0.03^{+0.02}_{-0.02}$&$-0.01^{+0.01}_{-0.01}$\\
II&$[\mathrm{Cu}/\mathrm{Fe}]$&$-0.41^{+0.19}_{-0.35}$&$-0.50^{+0.03}_{-0.03}$&$-$\\
II&$[\mathrm{Zn}/\mathrm{Fe}]$&$+0.29^{+0.10}_{-0.08}$&$+0.13^{+0.02}_{-0.02}$&$-$\\
    \end{tabular}
    \label{tab:results_abundances}
\end{table}
\subsection{Results}
Fig.~\ref{fig:ns_results}, Fig.~\ref{fig:galah} and Fig.~\ref{fig:apogee} show the results of fitting these models to the \citetalias{NS10}, \citetalias{Buder2020} and \citetalias{SDSSDR16} \gs samples respectively. In Table~\ref{tab:results_parameters} we give the model parameters governing the star formation, outflow and delay-time distribution of the Type Ia supernovae, and in Table~\ref{tab:results_abundances} we give the abundances of the Type II and Type Ia channels. Here we discuss the results in general terms, mainly focusing on differences between the results from the different datasets. The next section concerns the meaning of the results in the context of nucleosynthesis in \ia explosions.

\begin{figure}
    \centering
    \includegraphics[width=\columnwidth]{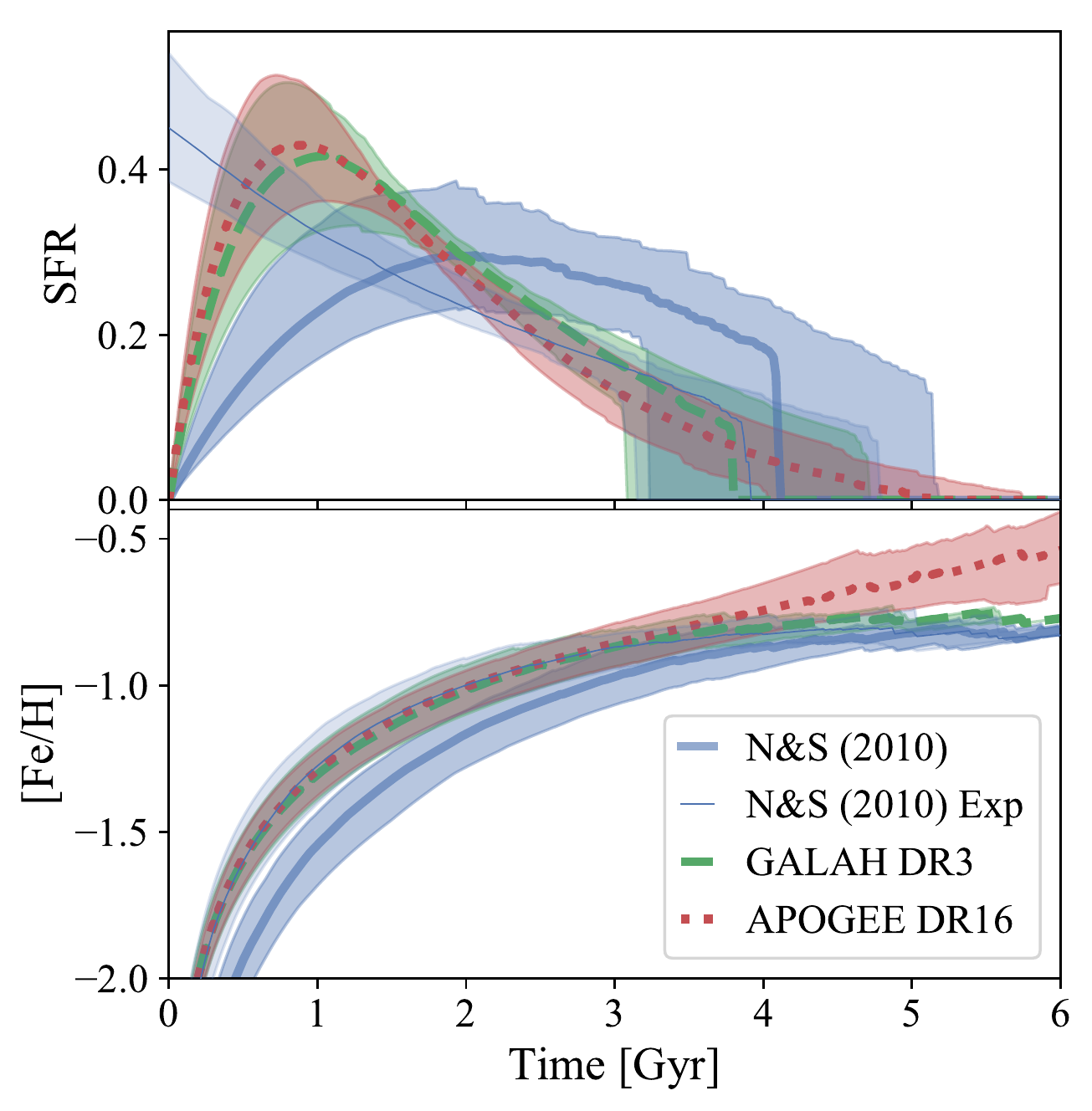}
    \caption{Comparison of the star formation history (top) and iron abundance over time (bottom) for the three \gs datasets: \protect\citet{NS10} in blue solid (with the exponential model variant shown with a thinner line), GALAH DR3 \protect\citep{Buder2020} in green dashed and APOGEE DR16 \protect\citep{SDSSDR16} in red dotted.}
    \label{fig::sfr_comparison}
\end{figure}

In the \citetalias{NS10} data we see little evidence for a plateau or knee in the $\xfemath{\alpha}$ abundances for the \gs stars suggesting the Type Ia supernovae began contributing below $\fehmath=-1.5\dex$. This is somewhat corroborated by the \citetalias{SDSSDR16} data where the knee occurs around $\fehmath\approx-1.5\dex$. However, the picture is less clear from the \citetalias{Buder2020} data where there is seemingly a plateau until a metallicity of $\fehmath\approx-1\dex$ and the subsequent turnover due to the Type Ia supernovae is less clear. All the studied $\alpha$ elements from \citetalias{Buder2020} have 1D non-LTE corrections which may explain the differing behaviours. 
We can see in Table~\ref{tab:results_parameters} that all models have a large $\tstar$ indicating inefficient star formation (high gas fraction) and high $\eta$ indicating a large outflow. These features are necessary to explain the down-turning $\xfemath{\alpha}$ at low metallicity. In Fig.~\ref{fig::sfr_comparison} we show the star formation histories and iron enrichment over time are very similar despite the differences in parameters. The \citetalias{NS10} data are equally well fitted using the exponential star formation history which matches better the constraints from the \citetalias{Buder2020} and \citetalias{SDSSDR16} data. Therefore, we conclude that the model parameter differences in Table~\ref{tab:results_parameters} lead to moderate changes in the shapes of the tracks. More important is the variation in the abundances ($m^\mathrm{Y}$) which alters the slope of the tracks. 

From Table~\ref{tab:results_abundances} we see that all models produce $\sim0.0015M_\odot$ per solar mass of star formation, consistent with the prior of $0.0017M_\odot$. On the other hand, the results for the iron contribution of Type II are slightly more varied with, in general, a larger contribution of $\sim0.0025M_\odot$ per solar mass of star formation compared to the prior of around $\sim0.00075M_\odot$. This is driven by the shape of the metallicity distributions which may be subject to some selection effects, and is also sensitive to the specific choice of star formation history. However, it could also point towards the initial mass function of the \gs producing more supernovae per generation. The $\alpha$ production of the Type II supernovae corresponds to the observed plateaus in each dataset, and the differences between different datasets can largely be attributed to the observed discrepancies between common stars (due in part to NLTE effects). The Type Ia production of $\alpha$ elements is low for all datasets and elements. Mg is consistent with the prior, demonstrating our models are sufficiently flexible to reproduce the Mg trends without invoking significant Mg production. For all datasets we see Ca production is greater than Si production (relative to solar) but the different dataset constraints are not consistent within the errors. This demonstrates the challenge of measuring the significant depletion of these elemental ratios.

We now focus on the iron-peak elemental abundances produced by the two channels. 
For all three datasets the three iron peak elements \xfe{Cr}, \xfe{Mn} and \xfe{Ni} are all relatively flat with metallicity over the range where $\xfemath{\alpha}$ is being depleted. In the case of Cr, this constrains the Type Ia and Type II abundance ratios to be very similar. The Type Ia yield for \citetalias{Buder2020} and \citetalias{SDSSDR16} are similar and slightly sub-solar, whilst \citetalias{NS10} is essentially solar, consistent with the discrepancies in the abundances for common stars discussed in Section~\ref{sec:common}. For Mn there is a weak positive slope present in all three datasets but most clearly in \citetalias{NS10}. This indicates a low (sub-solar) Type Ia Mn enrichment of around $\xfemath{Mn}\sim-0.15\dex$. \xfe{Ni} has a negative slope with metallicity in all datasets (note that using the uncorrected GALAH DR3 Ni abundances we found a positive slope). As with \xfe{Mn}, this points to a sub-solar \xfe{Ni} contribution from Type Ia. For both Mn and Ni, the measurements from all three datasets are consistent within the errors.

Cu and Zn are also produced predominantly in supernovae but with very weak contributions from AGB stars \citep[e.g.][]{KobayashiChemicalEvolution}. Additionally, as an odd-$Z$ element, Cu production (particularly from Type II supernovae) appears sensitive to metallicity \citep{KobayashiChemicalEvolution}. The model fits do not completely capture the trends in the GALAH data, possibly for this reason, although for Zn in particular the scatter is large. Despite these caveats, our results indicate a typical \xfe{Cu} from Type Ia of $\sim-0.5\dex$. The downwards trend in \xfe{Zn} from the \citetalias{NS10} data gives \xfe{Zn} $\sim-1\dex$, whilst the \citetalias{Buder2020} produces a result around solar (although the results are marginally consistent within the errors).

\begin{figure*}
    \centering
    \includegraphics[width=\textwidth]{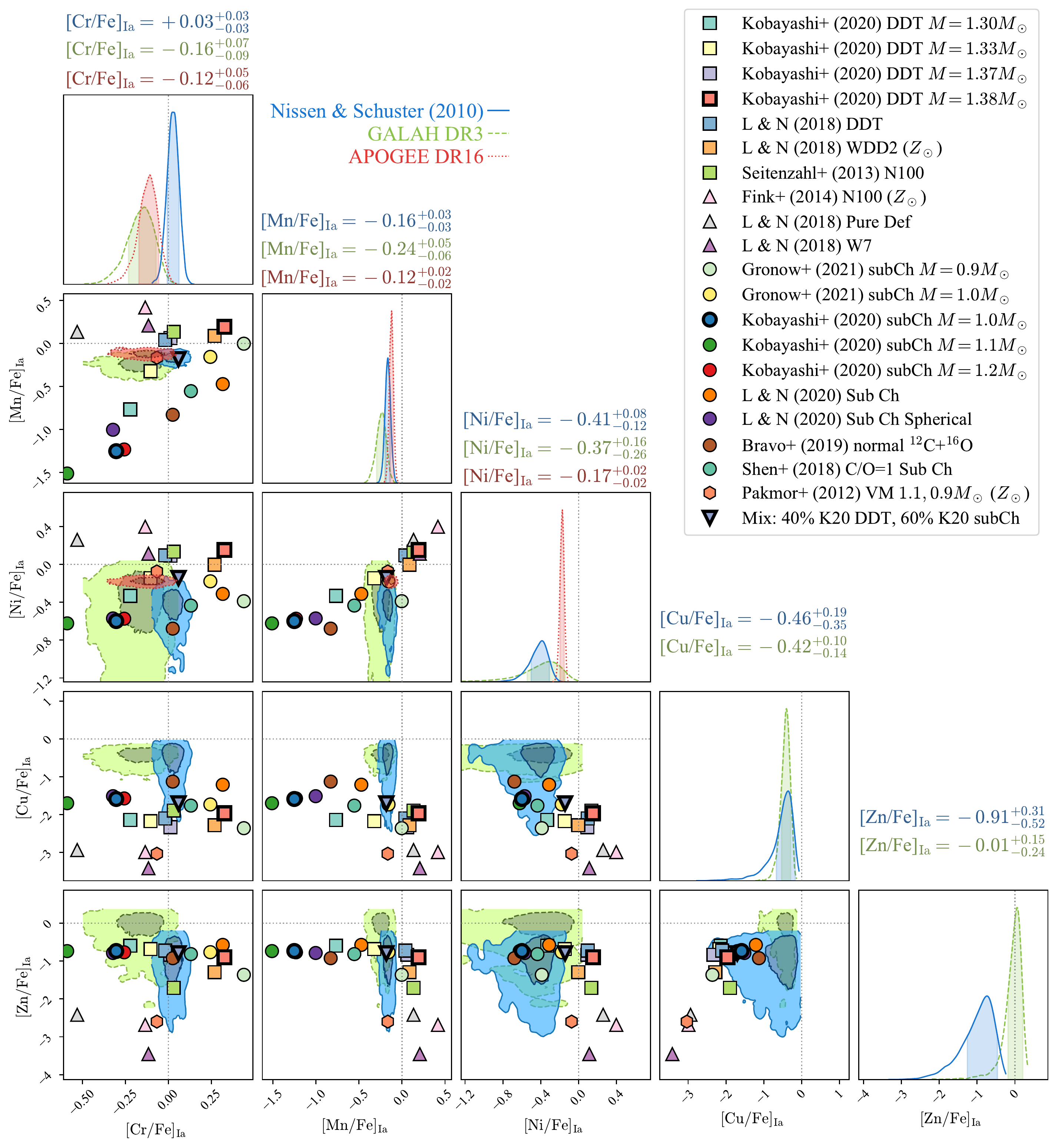}
    \caption{Constraints on the elemental yields from Type Ia supernovae determined from the \protect\citetalias{NS10} (blue solid), \protect\citetalias{Buder2020} (green dashed) and \protect\citetalias{SDSSDR16} (red dotted) \gs samples compared to theoretical yields from the literature. Squares denote deflagration-to-detonation (DDT) scenarios, triangles are pure deflagrations (Def) and circles are sub-Chandrasekhar mass double-detonation Type Ia models (see Section~\ref{sec::theoretical_models} for a detailed discussion of the different models). Although no model set exactly reproduces all of the data, note how the sub-Chandrasekhar mass models produce lower quantities of Mn and Ni consistent with the observations. Vertical and horizontal dashed lines are the solar values.}
    \label{fig:yields_corner}
\end{figure*}

\begin{figure}
    \centering
    \includegraphics[width=\columnwidth]{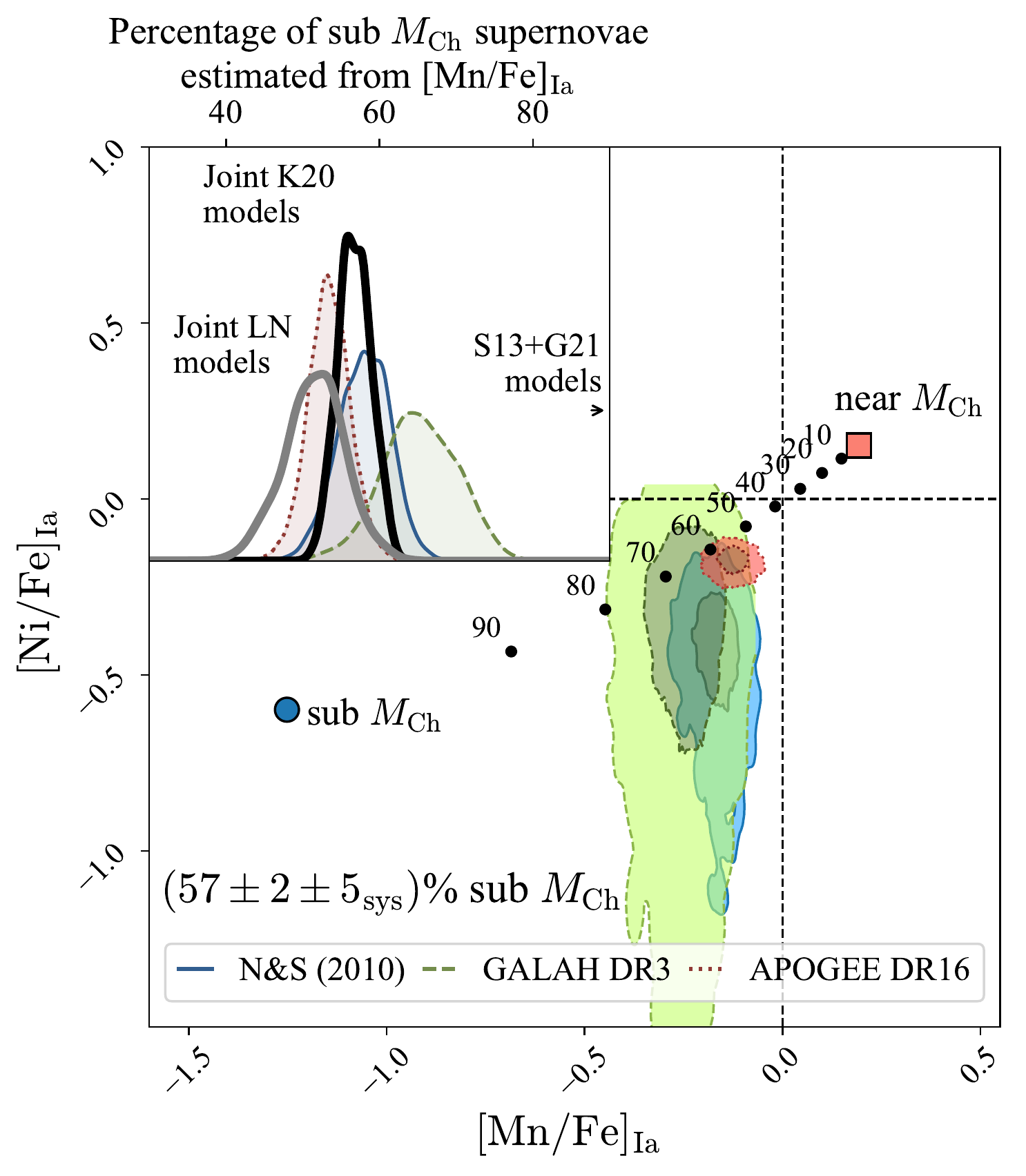}
    \caption{Constraints on the nickel and manganese yields from Type Ia supernovae determined from \protect\citetalias{NS10} (blue solid), \protect\citetalias{Buder2020} (green dashed) and \protect\citetalias{SDSSDR16} (red dotted) \gs samples. A \Ch deflagration-to-detonation transition model and a \sCh double-detonation model are shown by the coloured points \citep[both from][]{Kobayashi2020} and the sequence of black points show the different ratio combinations labelled by the percentage of \sCh supernovae. The inset shows the inferred percentage of sub-Chandrasekhar Type Ia supernovae required to reproduce the Type Ia \xfe{Mn} measurements. The coloured pdfs show the individual sample results assuming the models of \protect\cite{Kobayashi2020}, whilst the black pdf shows the combined result. The grey pdf shows the combined result assuming the models of \protect\cite{LeungNomoto2018,LeungNomoto2020} which adopt a different initial composition. The arrow shows the $95\,\percent$ lower bound using the models of \protect\cite{Seitenzahl2013} and \protect\cite{Gronow2021a,Gronow2021b}.}
    \label{fig:mn_ni_alt}
\end{figure}

\begin{figure*}
    \centering
    \includegraphics[width=\textwidth]{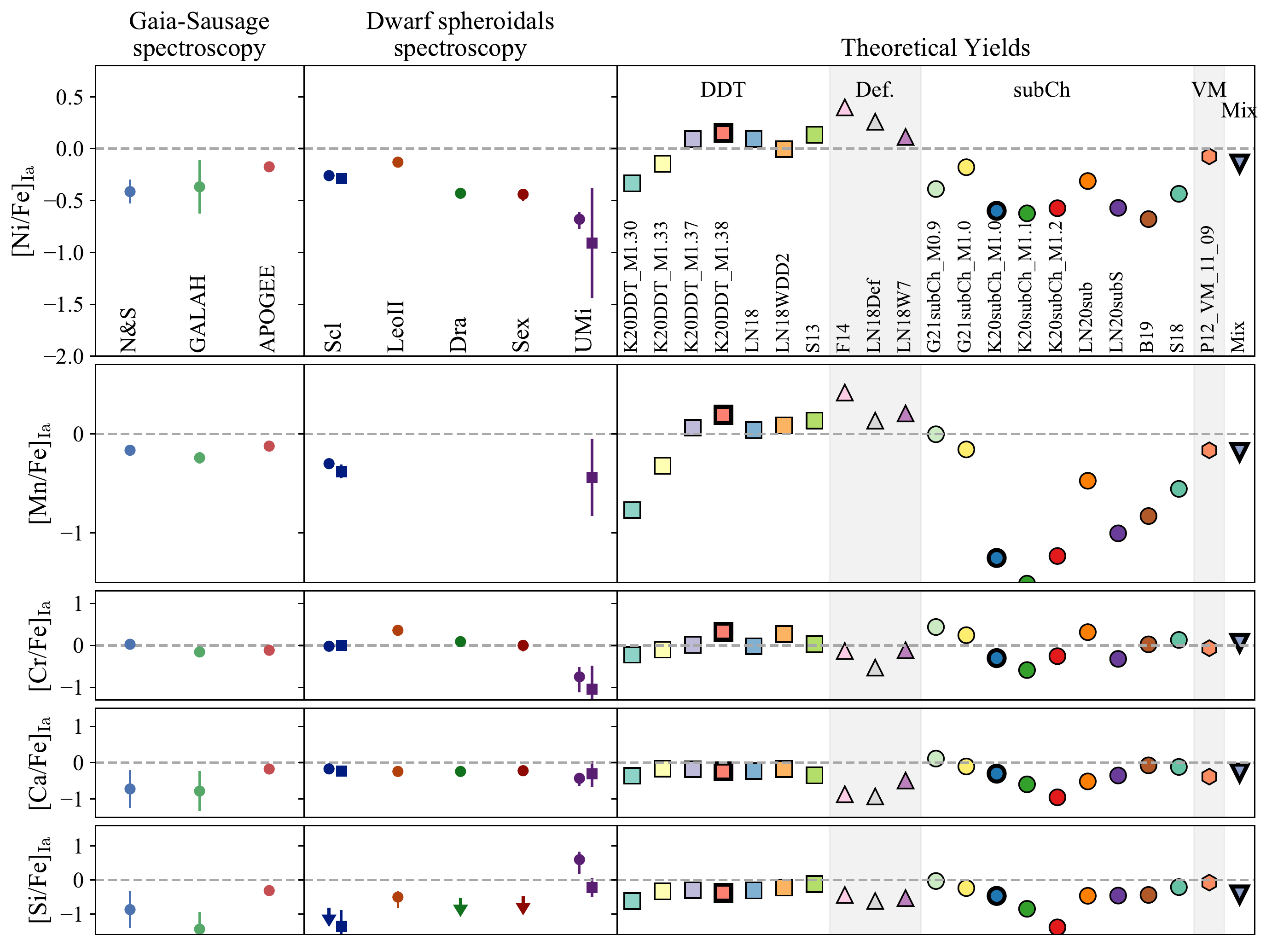}
    \caption{Comparison between the Type Ia supernovae yields in the \gs compared with (i) Milky Way dwarf spheroidal galaxies as measured by \protect\cite{Kirby2019} and \protect\cite{delosReyes2020} for Mn (circles) and the model fits from Section~\ref{sec::metdepyields} (squares), and (ii) theoretical Type Ia yields as defined in Table~\ref{tab:models_offsets} (markers with thick black outlines show benchmark models which reproduce observed $^{56}$Ni masses and the adopted mixture of the benchmark models). Sub-solar \xfe{Mn} and \xfe{Ni} are observed in both the \gs and the dwarf spheroidals, indicating the importance of sub-Chandrasekhar mass Type Ia supernovae in low mass/low metallicity systems.
    }
    \label{fig:dsph}
\end{figure*} 

As a check of the robustness of the results, we also considered the exponential star formation law solution from \cite{Weinberg2017} and found the model fits and the conclusions regarding the inferred yields very similar (see Table~\ref{tab::model_difference} for a comparison for the \citetalias{NS10} dataset). 

\section{Discussion}\label{sec::discussion}
We now discuss the constraints derived from the chemical evolution models of the \gs with comparison to theoretical calculations of Type Ia supernovae yields and similar constraints for dwarf spheroidal galaxy data.

\subsection{Theoretical models}\label{sec::theoretical_models}
As discussed in the Introduction, we consider three main categories of theoretical Type Ia supernovae explosion calculation: deflagration-to-detonation transition, pure deflagration and double-detonation models. All considered models (bar one) follow a single white dwarf system which is triggered to explode in a variety of ways. 
We evaluate the theoretical yields for a SN Ia with metallicity $Z=10^{-1.2}Z_\odot$ by linearly interpolating in $\log_{10}(Z)$ or if $Z=10^{-1.2}Z_\odot$ is not covered by the grids we use the lowest available metallicity model. Only the pure deflagration WDD2 model of \citet{LeungNomoto2018}, the pure deflagration model of \citet{Fink2014} and the violent merger model of \citet{Pakmor2012} are only available at solar metallicity.

\subsubsection{Deflagration-to-detonation transition models}
    Deflagration-to-detonation transition (DDT) models are the favoured types of models for \Ch \ia. A sub-sonic deflagration phase is introduced in a carbon-oxygen white dwarf which initially burns at high density producing iron-group elements. This causes the star to expand producing lower density regions containing unburnt fuel. After $\sim1\,\mathrm{s}$ the transition occurs to a super-sonic detonation phase which burns the remaining fuel predominantly producing intermediate mass elements (such as Si) due to the now lower densities. The deflagration flame front is highly textured in this scenario so multi-dimensional models are essential. We consider three different sets of models. \citet[][S13]{Seitenzahl2013} present yields from 3D DDT models using a varying number of ignition sites and four metallicities ($Z=Z_\odot,0.5Z_\odot,0.1Z_\odot,0.01Z_\odot$). We use the benchmark models with $100$ ignition sites and a central density of $2.9\times10^9\,\mathrm{g}\,\mathrm{cm}^{-3}$ which produce $\sim0.6M_\odot$ of $^{56}$Ni. The strength of the deflagration phase increases with the number of ignition sites, which decreases the production of Fe, Cu and Zn but increases the production of intermediate mass elements due to increased energy injection and expansion and increases the production of Cr, Mn and Ni. \citet[][LN18]{LeungNomoto2018} ran a series of 2D DDT models with a range of central densities, metallicities, C/O ratios and ignition kernels. We consider their benchmark models with central density $3\times10^9\,\mathrm{g\,cm}^{-3}$ at $Z = 0,0.1,0.5 ,1,2,3,5Z_\odot$ which produce $0.63M_\odot$ of $^{56}$Ni, and their updated WDD2 model \citep{Iwamoto1999} which we denote LN18WDD2. \cite{Kobayashi2020} updated the models of \cite{LeungNomoto2018} using a more realistic solar-scaled initial composition (elements lighter than Ne are presumed to be converted into Ne during normal H and He burning), again varying both the initial mass and metallicity ($Z=0$ to $Z=0.04$). As in \cite{LeungNomoto2018}, the \Ch model with a central density of $3\times10^9\,\mathrm{g}\,\mathrm{cm}^{-3}$ produced $0.63M_\odot$ of $^{56}$Ni. Lower masses produce more Fe and less Si, Ca, Cr, Mn, Ni and Zn. We consider all four mass models ($1.30,1.33,1.37,1.38M_\odot$). We denote the models as K20DDT\_M*. The lower of these masses can be considered as \sCh models. However, the difference with the double-detonation scenario (see below) is that central carbon ignition does not initiate a detonation due to the high electron-degeneracy pressure, but rather a deflagration. Also, in these cases, the physical trigger for the explosion isn't clear. Prior to ignition, a near-\Ch WD undergoes a simmering period where low-level carbon burning slowly increases the neutron fraction. This is equivalent to raising the metallicity of the WD, such that below a metallicity floor of $Z/Z_\odot\approx2/3$ \citep{PiroBildsten2008}. The models of \cite{Kobayashi2020} and \cite{Seitenzahl2013} have relatively weak metallicity dependence on the Mn production so we ignore this detail.
    
    \subsubsection{Pure deflagration models}
    Pure deflagrations (Def) have been proposed as scenarios to explain Type Iax supernovae \citep[spectroscopically similar to Type Ia but with lower peak brightness,][]{Foley2013}. The deflagration does not completely destroy the star and they primarily produce intermediate-mass elements with lower iron yields and an overabundance of neutron-rich species (e.g. $^{58}$Ni). Their contribution to galactic chemical evolution is believed to be quite small \citep{KobayashiChemicalEvolution}. \citet[][F14]{Fink2014} ran 3D pure deflagration models using a similar setup to \cite{Seitenzahl2013}. We select the model with $100$ detonation sites. Mn production has a weak dependence ($\sim0.2\dex$ variation) on the number of ignition sites as shown in \cite{delosReyes2020}. We also take the 2D pure deflagration models from \citet[][LN18Def]{LeungNomoto2018}. \cite{LeungNomoto2018} have rerun the classic W7 model of \cite{Iwamoto1999} using updated electron capture rates and run a series of new models with varying central density. We take the model with a central density of $10^9\,\mathrm{g\,cm}^{-3}$ (LN18W7). Higher central densities lead to higher Mn yields.
    
    \subsubsection{Double-detonation models}
    Double-detonation (DD) models are the favoured scenario for \sCh \ia. In this scenario, He is accreted onto the surface of the CO WD from a non-degenerate He-star, a companion He WD or even another CO WD as they are expected to have a small surface He layer. The He shell ignites and detonates at the base of the layer when it becomes sufficiently massive or if driven dynamically through the violent merger of two white dwarfs \citep{Pakmor2012}. This initial detonation triggers a subsequent detonation in the C/O core. There are two general classes of these controlled detonation models: those with and without a He shell. The He shell is required to initiate the first detonation but as observations suggest only a low mass He shell is required \citep{Kromer2010}, several groups have run pure detonation simulations without a He shell \citep[e.g.][]{Sim2010}. Both \citet[][S18]{Shen2018} and \citet[][B19]{Bravo2019} ran 1D bare detonation simulations. \cite{Bravo2019} considered five WD masses at five metallicities at two $^{12}$C+$^{16}$O reaction rates (standard and reduced by a factor 10). Here we use the $1.06M_\odot$ model with standard $^{12}$C+$^{16}$O reaction rate. Similarly, \cite{Shen2018} ran 1D bare detonation models at different masses, metallicities, C/O compositions and  $^{12}$C+$^{16}$O reaction rates. We use the $1M_\odot$ model with C/O=50:50 and standard $^{12}$C+$^{16}$O reaction rate. As discussed by \citet{Gronow2021a} for example, although of low mass, the He shell is important to the explosion mechanism and subsequent yields of the model. \cite{LeungNomoto2020} ran a series of 2D double-detonation simulations varying the core mass, the He shell mass, metallicity and initial He detonation mechanism. We consider their benchmark model with core mass $M=1.1M_\odot$, a $0.1M_\odot$ He shell and detonated with a bubble 50km above the core-envelope interface (110-100-x-50) which produces $0.62M_\odot$ of $^{56}$Ni. We also consider the benchmark models with a spherical detonation which have $M=1.0M_\odot$ and a $0.05M_\odot$ He shell (100-050-x-S50). Differences in the detonation geometry are largely characterised by differences in the abundances of light iron-peak elements (e.g V and Cr) with aspherical detonations producing higher abundances of these elements \citep[][although note that the two inspected models differ in their He shell mass which also affects these elements]{LeungNomoto2020,Palla2021}. As with the DDT models, the models of \cite{LeungNomoto2020} were updated with a more realistic initial mix of heavy elements by \cite{Kobayashi2020} who ran models at a series of core masses and metallicities ($Z=0$ to $Z=0.04$) with $0.05M_\odot$ He shell and a spherical detonation pattern. We use the $1,1.1$ and $1.2M_\odot$ total mass models, ignoring the $0.9M_\odot$ model as it produces a low quantity of Fe. The $1M_\odot$ model produces $0.63M_\odot$ of $^{56}$Ni. Very recently, \citet[][G21]{Gronow2021a,Gronow2021b} have presented a series of 3D double-detonation models investigating a range of core masses and He shell masses, and post-processed for a range of metallicities ($Z=0.01Z_\odot,0.1Z_\odot,Z_\odot,3Z_\odot$). We use the models with $0.9$ and $1M_\odot$ core mass and $0.05M_\odot$ He shell which produce around $0.6M_\odot$ of $^{56}$Ni, and ignore the $0.8M_\odot$ model as the iron production is low. Despite similar core and shell masses, the \cite{Gronow2021a,Gronow2021b} model yields differ significantly from those of \cite{Kobayashi2020} possibly related to the core-shell mixing in \cite{Gronow2021a,Gronow2021b}, the difference in the dimensionality of the models, or the difference in the nuclear network size (7 isotopes in \citealp{Kobayashi2020} compared to 35 in \citealp{Gronow2021a,Gronow2021b}) as for the low density regions found in \sCh WDs some burning pathways are neglected by smaller networks \citep{Shen2018}.
    
    The described bare and double-detonation models all follow the explosion of a single white dwarf presuming some initial configuration from the result of an ongoing merger event. Simulations of the violent mergers of two white dwarfs \citep{Pakmor2012} have been considered but there are more limited available nucleosynthetic yields for this case due to the increased complexity of the simulations. A set of four violent merger models from \cite{Pakmor2010, Pakmor2012} and \cite{Kromer2013, Kromer2016} are available through the HESMA database\footnote{\href{https://hesma.h-its.org}{https://hesma.h-its.org}}. Of these, only the model of \citet[][P12\_VM\_11\_09]{Pakmor2012} of a $1.1$ and $0.9M_\odot$ WD merger produces a reasonable quantity of Fe ($\sim0.6M_\odot$) so we ignore the others. This model is essentially a pure/bare detonation model. As \cite{Seitenzahl2013B} reports, the model produces $\xfemath{Mn}=-0.15\dex$.

\subsubsection{Comparison with observations}
Fig.~\ref{fig:yields_corner} shows the Type Ia abundance constraints from the \gs datasets with the theoretical yields from the different models. We show the five iron-peak elements in our modelling (Cr, Mn, Ni, Cu and Zn). Focusing on Mn and Ni, we see that all models lie along an anti-correlation in $\xfemath{Mn}_\mathrm{Ia}$ vs. $\xfemath{Ni}_\mathrm{Ia}$ as highlighted by \cite{Kobayashi2020}. As discussed in the Introduction, these elements are particularly sensitive to the density and hence mass of the model. Pure deflagration models sit at the high abundance end of the trend along with the DDT models. The lower mass DDT models of \cite{Kobayashi2020} sit further down the trend with the \sCh models as they too can be considered as \sCh models. The violent merger model of \cite{Pakmor2012} sits part way between the two groups of models. It is clear our derived constraints are able to distinguish between the different scenarios. Based on this panel alone, all datasets are consistent with either \begin{inparaenum}
\item the $1M_\odot$ G21 model,
\item a mixture of `normal' DDT (40\%) and \sCh (60\%) \cite{Kobayashi2020} models (as shown in the plot),
\item the violent merger model of \cite{Pakmor2012} or
\item a low mass $1.33M_\odot$ DDT model (which is perhaps not well physically motivated).
\end{inparaenum}
The distribution of the models in Cr abundances is not so easy to understand and is perhaps more reflective of the ignition mechanism rather than its structure \citep{Palla2021}. There are DDT models and \sCh models consistent with the solar/sub-solar measured abundances. The mixture model matches the \citetalias{NS10} well and as discussed these are systematically higher than the corresponding \citetalias{Buder2020} and \citetalias{SDSSDR16} models. Finally, Cu and Zn are heavier iron-peak elements largely produced in the detonation phase \citep[e.g.][]{Kobayashi2020}. We observe that deflagration models are inconsistent with the constraints. Also, despite its success in reproducing \xfe{Mn} and \xfe{Ni}, the violent merger model underproduces both of these elements. \sCh models tend to produce more Cu than DDT models producing better consistency with the data. The Zn measurements of essentially all DDT and \sCh models are consistent with the data.

In Fig.~\ref{fig:mn_ni_alt} we show a larger version of the $\xfemath{Mn}_\mathrm{Ia}$ vs. $\xfemath{Ni}_\mathrm{Ia}$ panel of Fig.~\ref{fig:yields_corner} with only the two benchmark \Ch and \sCh models of \cite{Kobayashi2020} shown. We also display the sequence of linear combinations of these two models. The \citetalias{SDSSDR16} constraints agree very well with a mixture of $60\,\percent$ \sCh \ia. Both the \citetalias{NS10} and \citetalias{Buder2020} are less conclusive in this 2D projection although they are consistent at the $2\sigma$ level with the APOGEE result. The $\xfemath{Mn}_\mathrm{Ia}$ measurement is more robust as for \citetalias{Buder2020} $\xfemath{Ni}$ we adjusted it based on overlaps with APOGEE. Also $\xfemath{Ni}_\mathrm{Ia}<\xfemath{Ni}_\mathrm{II}$ whilst $\xfemath{Mn}_\mathrm{Ia}>\xfemath{Mn}_\mathrm{II}$ meaning we can be more confident about the Mn contribution from \ia. In the inset panel of Fig.~\ref{fig:mn_ni_alt} we display the inferred \sCh percentage based on the Mn measurements alone. \citetalias{Buder2020} yields slightly higher percentages of $\sim65\,\percent$ whilst \citetalias{NS10} and \citetalias{SDSSDR16} are more consistent with $\sim55\,\percent$. This is despite there being minimal differences in the Mn abundances from common stars so the difference must be driven by how the distribution of all the abundances constrains the \ia contribution. Combining the constraints gives $(57\pm2)\,\percent$ \sCh. Adopting instead the \cite{Kobayashi2020} $1.1$ or $1.2M_\odot$ total mass \sCh models yields essentially identical constraints of $56$ and $57\,\percent$ respectively so any mass distribution of \sCh WDs would produce a similar constraint.

If instead we adopt the models of \cite{LeungNomoto2018} (which include only $^{22}$Ne for the metallicity rather than a scaled-solar composition and so are perhaps less realistic) we yield $\sim52\,\percent$ \sCh using the $1.1M_\odot$ core mass \sCh model and $\sim40\,\percent$ using the spherical detonation $1.0M_\odot$ model. The more recent metallicity-dependent 3D \sCh models from \cite{Gronow2021a,Gronow2021b} in general produce significantly higher \xfe{Mn} than the \cite{Kobayashi2020} models. The $1M_\odot$ core mass model implies essentially $100\,\percent$ \sCh supernovae ($\gtrsim87\,\percent$ \sCh \ia at $95\,\percent$ confidence), whilst the $1.1M_\odot$ core mass model requires $(67\pm3)\percent$ \sCh \citep[if combined with the][models]{Seitenzahl2013} and the $0.9M_\odot$ produces too high \xfe{Mn} to be consistent with the data. Although our inference is still quite uncertain as it depends on the specific models we adopt, 
the general conclusion of a significant fraction of \sCh appears robust.

\subsection{Non-LTE effects}\label{sec::nlte}

Mn abundance measurements are known to be sensitive to the assumption of local thermal equilibrium \citep[LTE,][]{BergemannGehren2008,Bergemann2019}. As our conclusions regarding the \ia enrichment channel depend on Mn, it is worth discussing the status of non-LTE (NLTE) corrections for Mn. In general, the difference between 1D NLTE Mn abundances and 1D LTE Mn abundances ($\Delta_\mathrm{NLTE}$) is positive, increases for lower metallicity stars and is typically larger for giant stars than dwarf stars, although the differences are line dependent with resonance lines more sensitive to NLTE effects. The metallicity dependence has the effect of steepening correlations in \xfe{Mn} vs. \feh when using LTE abundances. \cite{BattistiniBensby2015} used the corrections from \cite{BergemannGehren2008} to produce NLTE Mn abundances for a sample of G dwarfs finding a very flat, near-solar \xfe{Mn} vs. \feh distribution for $-1\lesssim\fehmath\lesssim0.5$. However, the updated NLTE corrections from \cite{Bergemann2019} suggest a shallower trend of $\Delta_\mathrm{NLTE}$ with metallicity. \cite{Eitner2019} use the updated model to compute (line-dependent) corrections for $\fehmath=-2\,\dex$ models finding $\Delta_\mathrm{NLTE}=0.2-0.25\,\dex$ for dwarf stars and $\Delta_\mathrm{NLTE}=0.35-0.4\,\dex$ for giants. \cite{Eitner2020} applied the new corrections to dwarf and giant stars between $-3\lesssim\fehmath\lesssim0$ finding a slowly rising \xfe{Mn} with \feh from $\xfemath{Mn}\approx-0.2\,\dex$ at $\fehmath\approx-2\,\dex$ to $\xfemath{Mn}\approx0\,\dex$ at $\fehmath\approx0\,\dex$. Five stars in the \cite{Eitner2020} sample are in the \cite{NS10} sample. Ignoring the measurement for HD 121004 which is based upon a Mn II line, we find the difference between the \cite{Eitner2020} NLTE measurements and the \cite{NS11} LTE measurements is $\sim0.07\,\dex$ but they do not show a significant increase with decreasing metallicity.

The \cite{NS11} Mn measurements are based upon Mn I lines at $4783.4, 4823.5, 6013.5, 6016.7$ and
$6021.8$\AA. The lines around $6015$\AA\ are weaker and not detectable at low ($\sim-1.4\dex$) metallicity. However, the lines around $4800$\AA\ are strong but ignored in cool, metal-rich stars (they can also lie outside the UVES spectral range). Assuming the dwarf model from \cite{Bergemann2019} both of these sets of lines have $\Delta_\mathrm{NLTE}=0.04-0.105\fehmath$. This can be compared to the correction in Figure 12 of \cite{Amarsi2020} which gives $\Delta_\mathrm{NLTE}\approx-0.1\fehmath$ (independent of whether the star is a dwarf or giant although \cite{Bergemann2019} suggest larger corrections for giants: $\Delta_\mathrm{NLTE}\approx0.2-0.1\fehmath$, but they do not provide results for the GALAH lines $4754.04$\AA\ and $4761.51$\AA). We have applied the \cite{Bergemann2019} dwarf NLTE correction to the \citetalias{NS10} data and refitted the chemical evolution models, despite the direct comparison of the four stars in \cite{Eitner2019} suggesting this correction is too large. This fit produces $\xfemath{Mn}_\mathrm{Ia}=(-0.17\pm0.04)\dex$, near identical to the results using the 1D LTE. The bigger difference is in the Type II yields which are $\sim0.2\,\dex$ higher at $\xfemath{Mn}_\mathrm{II}=(-0.18\pm0.05)\dex$. This higher Mn production could be matched by the yields of \cite{ChieffiLimongi2004} or the fast rotating models of \cite{LimongiChieffi2018}. The \citetalias{Buder2020} abundances already include the 1D NLTE corrections for Mn yet yield a higher $\xfemath{Mn}_\mathrm{Ia}$ than the corrected \citetalias{NS10} data. It should also be noted that similar trends in \xfe{Mn} vs. \feh for the \citetalias{NS10} dataset are seen in both the \gs and the higher-$\alpha$ in-situ or \emph{Splash} \citep{Belokurov2020} stars. As noted in \cite{NS11}, this does suggest some part of the trend is due to NLTE effects although clearly the two groups are still distinct in this projection.

The 1D NLTE corrections do not tell the full story, as 3D corrections are also important for Mn. However, 3D calculations are expensive so much more limited information is available. \cite{Bergemann2019} compute the difference between 1D and 3D NLTE calculations for a set of four models (giant and dwarf at $\fehmath=-1,-2\,\dex$) finding, in general, 3D NLTE Mn abundances are larger than 1D NLTE. 
From \cite{Bergemann2019} the lines around $4800$\AA\ and $6015$\AA\ in dwarf stars have 3D NLTE larger than 1D by $0.06\,\dex$ and $0.08\,\dex$ respectively. Applying this correction to our dataset will shift the measurement of $\xfemath{Mn}_\mathrm{Ia}$ up by $\sim0.05\dex$ which would lower the contribution of sub-\Ch to around $50\,\percent$ in line with the Milky Way result from \cite{Seitenzahl2013B}.

GALAH uses the lines at $4754.04$\AA\ and $4761.51$\AA. 
which have 3D NLTE larger than 1D NLTE by approximately $0.02-0.05\,\dex$ and $0.08-0.09\,\dex$ for the $-2$ and $-1\,\dex$ stars respectively. Interestingly, for Mn I in the infra-red, \cite{Bergemann2019} report that 3D NLTE should be consistently higher than 1D NLTE by $0.15\,\dex$ for all analysed stars. However, as shown in Appendix~\ref{appendix::common_stars}, no difference between the 1D LTE APOGEE and the 1D NLTE GALAH Mn abundances is observed, whilst the results in \cite{Bergemann2019} suggest a possible discrepancy of $\sim0.15\dex$ at $-1\dex$ and $\sim0.2\dex$ at $-2\dex$ (APOGEE being lower).

Although there is clear evidence that 1D LTE Mn abundances are too low at low metallicity, NLTE corrections depend sensitively on the details of the modelling and input data. Their dependence on effective temperature and surface gravity can be tested by requiring globular cluster stars have similar abundances. For example, \cite{Kirby2018} applied Cr and Co NLTE corrections to LTE measurements for globular cluster members but found the corrections introduced systematic trends which were not present in the LTE calculations. However, \cite{Kovalev2019} show how NLTE measurements of Fe, Mg, and Ti globular cluster stars do not increase the intra-cluster scatter. These studies suggest that it is far from trivial to apply corrections to the LTE abundances.

\begin{figure*}
    \centering
    \includegraphics[width=\textwidth]{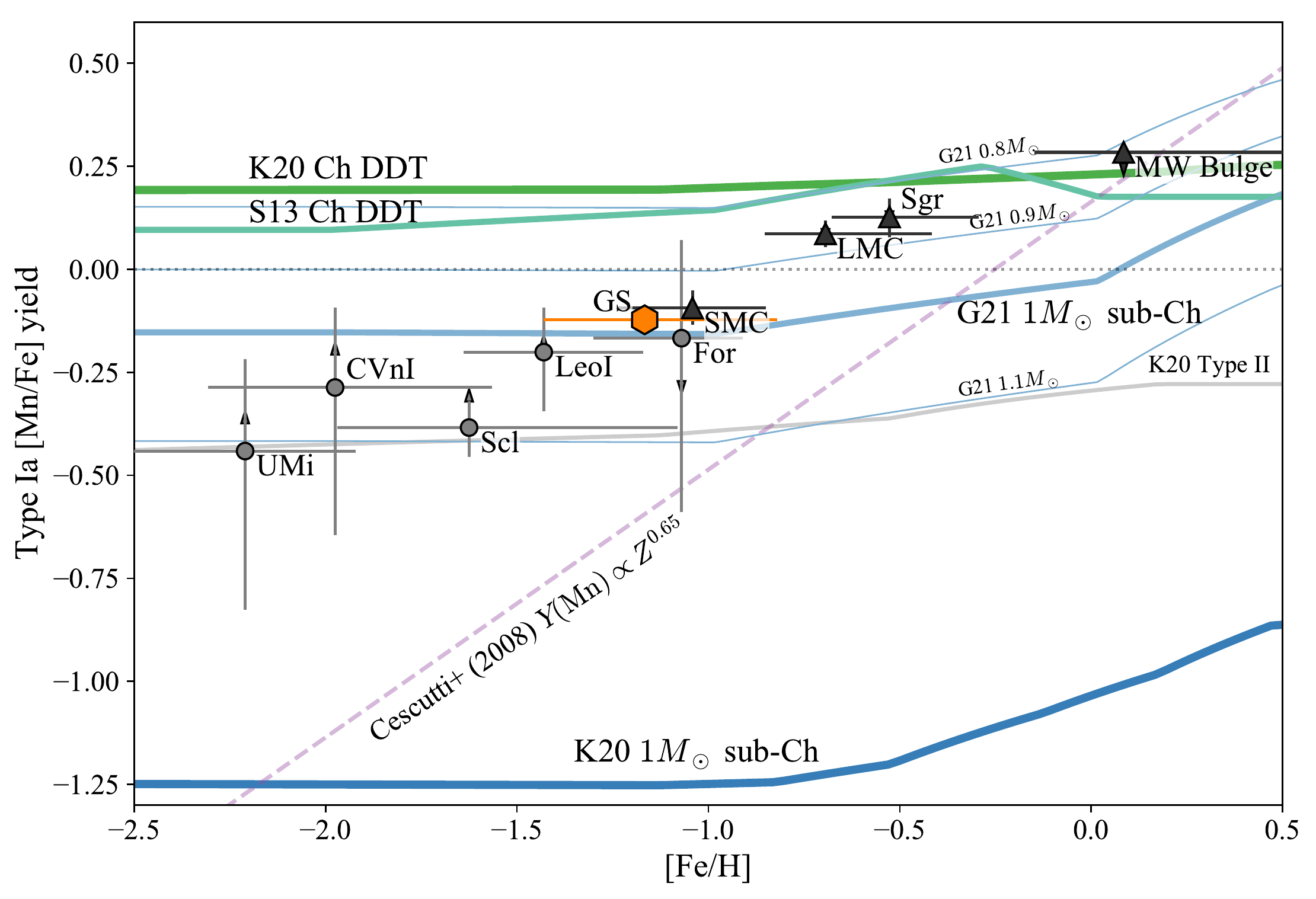}
    \caption{Inferred Type Ia $\xfemath{Mn}$ yield against metallicity for a range of environments. We show data for the Milky Way Bulge, the Large and Small Magellanic Clouds, the \gs (orange hexagon) and a number of dwarf spheroidal galaxies. Vertical errorbars show the uncertainty whilst horizontal errorbars give the metallicity spread of the considered stars (16th and 84th percentiles). The small arrows show the effect of correcting for NLTE effects by subtracting $0.1\fehmath$ from the individual abundances. We plot the \Ch deflagration-to-detonation metallicity-dependent yields from
    \protect{\citet[][S13 Ch DDT, dark green]{Seitenzahl2013}} 
    and \protect{\citet[][K20 Ch DDT, light green]{Kobayashi2020}}, and the \sCh double-detonation metallicity-dependent yields from 
    \protect{\citet[][G21 $1M_\odot$ sub-Ch, thick light blue, and other masses in thinner lines]{Gronow2021a,Gronow2021b}} 
    and \protect{\citet[][K21 $1M_\odot$ sub-Ch, dark blue]{Kobayashi2020}}. 
    We also display in grey the Type II yields from \protect\cite{Kobayashi2006} used in the \protect\cite{KobayashiChemicalEvolution} models. The purple dashed line is the proposed metallicity dependence of the Type Ia Mn yields from \protect\cite{Cescutti2008}.}
    \label{fig:met_dep}
\end{figure*}

\subsection{Comparison with Milky Way dwarf spheroidal galaxies}
In a similar vein to our work, \cite{Kirby2019} and \cite{delosReyes2020} have estimated the contributions of different Type Ia supernovae to the five dwarf spheroidal (dSph) galaxies: Sculptor, Leo II, Draco, Sextans and Ursa Minor (UMi). These isolated low-metallicity systems provide an alternative way to inspect chemical evolution of entire galaxies. \cite{Kirby2019} fitted linear models with a break to \xfe{X} vs. \feh data with X=(Mg, Si, Ca, Cr, Fe, Co, Ni) from which the yields of the supernovae could be computed. The advantage of this procedure is that the metallicity dependence of the supernovae yields can be more simply captured. However, unlike our procedure where there is a common set of evolutionary parameters (e.g. star formation, outflow), the fits for each abundance are independent. Based on the sub-solar \xfe{Ni} of Type Ia supernovae, \cite{Kirby2019} concluded there must be a significant contribution from sub-Chandrasekhar mass Type Ia supernovae in these systems. \cite{delosReyes2020} measured Mn abundances for Sculptor and fitted the same models as \cite{Kirby2019} to find $\xfemath{Mn}_\mathrm{Ia}=(-0.30\pm0.03)\dex$. These authors also suggest that systems with early concentrated star formation (such as Sculptor) have lower $\xfemath{Mn}_\mathrm{Ia}$, and hence higher fractions of sub-Chandrasekhar mass Type Ia supernovae, than systems with more extended star formation. This suggests that perhaps the dominant channel of Type Ia enrichment depends on delay time with Chandrasekhar mass systems dominating at late times.

In Fig.~\ref{fig:dsph} we show our constraints for the \gs alongside the constraints from the dSph models of \cite{Kirby2019} and \cite{delosReyes2020} and the theoretical models of the previous section. Sculptor has a sub-solar Type Ia \xfe{Mn} yield that is $\sim0.1\,\dex$ lower than that for the \gs. As for the \gs result, the Sculptor result is consistent with a mixture of \Ch and \sCh from the \cite{Kobayashi2020} models or near $100\,\percent$ \sCh from the \cite{Gronow2021b} or \cite{Pakmor2012} models. Similarly, both Sculptor or Leo II have sub-solar $\xfemath{Ni}_\mathrm{Ia}$ like the \gs. Draco and Sextans have lower $\xfemath{Ni}_\mathrm{Ia}$ still and Ursa Minor has the lowest $\xfemath{Ni}_\mathrm{Ia}$ consistent only with the \sCh yields. Cr presents a more complicated picture with all results (except that of UMi) around solar. This is consistent with almost any set of theoretical yields with the results of UMi suggesting a large fraction of \sCh or even pure deflagration models. The analysis from \cite{Palla2021} demonstrates that Cr is more sensitive to the details of the explosion initialization and geometry, and so is more an indicator of different simulation setups. The $\alpha$ elements (Si and Ca) are sub-solar for all dwarf spheroidals as well as for the \gs (again except for UMi which has a high \xfe{Si}). As with Cr, this is largely consistent with any set of theoretical \ia yields.

\subsection{Type Ia yields across a range of systems}\label{sec::metdepyields}
The comparison with the dwarf spheroidal data in the previous section demonstrates that both the \gs and dSphs have significant contribution from \sCh \ia. However, there is also the suggestion of metallicity dependence in the Type Ia yields with Sculptor having a $\sim0.1\,\dex$ lower \xfe{Mn} yield than \gs, despite both systems having similar early bursts of star formation. To investigate the potential metallicity dependence of the Type Ia yields, we fit the simple analytic models to data from a wider range of systems. We consider \citetalias{SDSSDR16} data for the Milky Way bulge (selected as giants within $5\,\mathrm{deg}$ of the Galactic centre), the Large Magellanic Cloud (LMC, fainter than $K_s=12$ within $15\,\mathrm{deg}$ of the centre of the LMC and with a line-of-sight velocity cut), the Small Magellanic Cloud (SMC, fainter than $K_s=12$ within $10\,\mathrm{deg}$ of the centre of the SMC and with a line-of-sight velocity cut) and the Sagittarius dSph (Sgr, fainter than $H=12$ within $5\,\mathrm{deg}$ of the centre of the Sgr, with a line-of-sight velocity cut and removing high-$\alpha$ in-situ stars). We use similar quality cuts on the abundances to those in Section~\ref{sec::data_apogee}. Furthermore we take the dSph data from \cite{Kirby2018} combined with the Mn measurements from \cite{delosReyes2020} for Ursa Minor, Canes Venatici I, Sculptor, Leo I and Fornax. 

We fit the analytic chemical evolution models to Mg, Si, Ca, Cr, Mn and Ni adopting identical priors to those in Section~\ref{sec::method} except for $t_\mathrm{max}$ and $\tau_\mathrm{sfh}$ which are chosen on the basis of star formation history measurements  \citep{deBoer_Sculptor,deBoer_Fornax,deBoer_Sgr,Barbuy,HarrisLMC,HarrisSMC,GallartLeoI,CarreraUrsaMinor}. We display the results of the model fits in Fig.~\ref{fig:met_dep} along with the theoretical yields from the \Ch models of \cite{Seitenzahl2013} and \cite{Kobayashi2020}, and the \sCh $1M_\odot$ models from \cite{Kobayashi2020} and \cite{Gronow2021a,Gronow2021b}. The full set of model constraints is given in Table~\ref{tab::lg_fullparams} and the model fits for the most massive systems are given in Fig.~\ref{fig:lg_model_comparison}. We see that for all systems of lower metallicity than the SMC there is a weak gradient of \xfe{Mn} with metallicity. This gradient may be consistent with NLTE effects. Based on the discussion of Section~\ref{sec::nlte}, we have rerun the modelling with a \xfe{Mn} NLTE correction of $-0.1\fehmath$, although this is possibly an overestimate of the effect. The shift in the results are shown by the small arrows in Fig.~\ref{fig:met_dep}. It is clear NLTE effects could flatten the gradient at the metal-poor end. However, more robust is the difference in the yield required to explain the higher metallicity systems (Milky Way bulge, LMC, Sgr) compared to the lower metallicity systems (e.g. \gs). The lower metallicity systems cover a range of star formation histories (e.g. the early star formation of \gs and Sculptor vs. the more extended star formation history from Fornax) as do the high metallicity systems (early star formation in the Milky Way bulge vs. extended, albeit bursty, star formation in the LMC). This suggests that although evolutionary differences may give rise to some variation at fixed metallicity \citep[see later and][]{delosReyes2020}, the main factor governing the Type Ia \xfe{Mn} yields is metallicity dependence. 

\subsubsection{Modelling variants}
Before discussing the implications of our results, we will detail some caveats and model variants we have explored to test the robustness of our derived constraints. 

\emph{Complex star-formation histories}: Several of these systems have more complicated star formation histories than can perhaps be captured by the linear-exponential star formation law of \cite{Weinberg2017}. For example, the Magellanic Clouds are known to have relatively extended, bursty star formation histories, possibly linked to their mutual interaction \citep{HarrisLMC,Nidever2020}, or the Milky Way bulge may have gone through bursts and quenching episodes \citep{Lian2020}. The models of \citet[][see section 5.6]{Weinberg2017} permit more complex star formation histories as linear sums of the analytic solutions (linear-exponential and exponential). Due to the linearity of the chemical evolution equations, the expressions for the \emph{mass} of each element are just linear sums of the solutions for each component. We have considered more complex models of two linear-exponential terms with relative weight $f_s$, with one star formation law beginning at $t=0$ and the other offset to begin at $t=t_0$, a free parameter. Each component has its own separate $\tsfh$. Despite the added complexity, the constraints on the yields, particularly Mn and Ni, are very similar to those obtained with the single component model. However, in general these models are much harder to sample from. Furthermore, removing the constraint that $t_i$ follows the star-formation law in the modelling produces significantly poorer matches to the metallicity distributions but the conclusions regarding $\xfemath{Mn}_\mathrm{Ia}$ and $\xfemath{Ni}_\mathrm{Ia}$ are unchanged. This suggests, despite the simplicity of the adopted star formation prescription and despite the sampling of the systems possibly being subject to selection effects that reshape the metallicity distributions, our results regarding the Mn and Ni abundances are robust to reasonable variations in the model star formation history parametrization. Note we are not arguing that the average Type Ia yields for each system do not correlate with the star formation history (see later discussion), rather our recovery of the average yields for each system is insensitive to modelling variations in the star formation history.

\emph{Metallicity-independent yields}: A further caveat is that several of the studied systems (in particular the Milky Way bulge) span a large range in metallicity such that the metallicity-independent yields prescription of the analytic models is possibly questionable. For example, in the case of the Milky Way bulge, the simplistic models fail to perfectly capture the morphology of the nearly linearly rising \xfe{Mn} vs. \feh. This is indicative of the need for metallicity-dependent yields for this system \citep{Gronow2021b}. For both of these reasons, our measurements should be considered as average \ia yields corresponding approximately to the metallicity at which the bulk of the stars were formed. 

\emph{Metallicity-dependent Type II yields} may also play a role and, as discussed, these are not well incorporated in our modelling framework. In Fig.~\ref{fig:met_dep}, we show the weak metallicity dependence of Type II yields from the models of \cite{Kobayashi2006,KobayashiChemicalEvolution} where the \xfe{Mn} difference between $\fehmath=-1$ and $0\,\dex$ is $\sim0.1\,\dex$. Other Type II supernovae yields give $\sim0.13\,\dex$ \citep{ChieffiLimongi2004}, $\sim0.27\,\dex$ \citep{WoosleyWeaver1995,Nugrid1,Nugrid2} or as large as $\sim0.6\,\dex$ \citep{LimongiChieffi2018,Prantzos2018} for this quantity. Such a large variation could in principle explain the \xfe{Mn} trend for the Milky Way bulge. However, the maximum $\xfemath{Mn}_\mathrm{II}$ yield at solar metallicity for any model is $\sim0.1\,\mathrm{dex}$ \citep{LimongiChieffi2018,Prantzos2018}. Assuming the same relative contribution to stars of solar metallicity from Type Ia and Type II yields as found in the default model fit, $\xfemath{Mn}_\mathrm{Ia}$ would be lowered from $0.28\,\mathrm{dex}$ to $\sim0.07\,\mathrm{dex}$ in this case. Even in this extreme case, super-solar Type Ia yields are required for high metallicity systems. As a further test, we incorporated a form of metallicity-dependent Type II yields in our models by including a different $\eta$ parameter for each element. As referenced in Section~\ref{sec::method}, a Type II yield, $m^\mathrm{II}_j$, which depends linearly on the current mass fraction of element $j$ relative to solar, $Z_j$, is equivalent to changing the depletion timescale $\tdep$ for this element. We included a unit normal prior on the gradient of $m^\mathrm{II}_j/m^\mathrm{II}_j(Z_j=0)$ with respect to $Z_j$ (consistent with the spread from the theoretical yields). Furthermore, we enforced the yield for a mass fraction $10$ times solar to be positive. This produced no difference in $\xfemath{Mn}_\mathrm{Ia}$ for systems with $\fehmath<-1\,\mathrm{dex}$ (including the \gs). For the LMC, Sgr and the Milky Way bulge, $\xfemath{Mn}_\mathrm{Ia}$ was significantly lower ($-0.09$, $-0.06$ and $-0.18\,\mathrm{dex}$ respectively) but, in the cases of the Milky Way bulge and the LMC, only by invoking unphysically steep Type II metallicity gradients such that $\xfemath{Mn}_\mathrm{II}\sim0.3-0.5\,\mathrm{dex}$ at solar metallicity. Thus it seems that although metallicity-dependent Type II yields could contribute to the interpretation of the data, they will only weakly flatten the gradient seen in Fig.~\ref{fig:met_dep} and the observation that the Milky Way bulge data require \ia Mn yields inconsistent with the lower metallicity systems (e.g. \gs) is robust. 

\emph{Tensions with priors}: From inspecting Table~\ref{tab::lg_fullparams}, there are a few parameter constraints which are in tension with our adopted priors. In particular, $\log_{10}m_\mathrm{Fe}$ for the Type II and Type Ia channels are in some instances much higher than the priors (e.g. the Milky Way bulge and the LMC). Running models with much tighter priors on these quantities ($0.004\,\mathrm{dex}$) produces very similar results for $\xfemath{X}_\mathrm{Ia,II}$ and by eye a similar quality of fit to the abundance trends. This is typically achieved through lowering $\tD$ such that it pushes against the $40\,\mathrm{Myr}$ lower bound and is consistent with very prompt \ia exploding soon after their progenitor star becomes a WD. Another quirk of our modelling is the low $\eta$ constraint for the Milky Way Bulge (consistent with zero) indicating very low outflow. Allowing $\eta$ to be negative produces a constraint of $\eta=(-0.15\pm0.12)$, which, as discussed previously, we could physically interpret as a Type II iron yield increasing with metallicity. Relaxing the positivity constraint on $\eta$ does not affect the inferred abundances.

We conclude that, despite some deficiencies in our modelling approach, the main results that we focus on are robust to reasonable variations of the model. This is further corroborated by the results of our models for Sculptor and Ursa Minor compared to the models of \cite{Kirby2019} and \cite{delosReyes2020} as shown in Fig.~\ref{fig:dsph}, demonstrating the good agreement between the two modelling frameworks. We now discuss the meaning of these results.

\subsubsection{Metallicity-dependent Type Ia yields}

Metallicity-dependent Type Ia \xfe{Mn} yields have long been proposed as an explanation for the rising \xfe{Mn} vs. \feh for Milky Way stars \citep[e.g.][]{Cescutti2008, Weinberg2019}. There are two possible routes for how such a metallicity dependence could arise. If we simply consider two different Type Ia channels, it could either be that the individual \xfe{Mn} yields from the two channels are metallicity dependent, or the relative contribution from the two channels is metallicity dependent. Both the models of \cite{Kobayashi2020} and \cite{Gronow2021b} demonstrate that metallicity dependence of $\xfemath{Mn}_\mathrm{Ia}$ can be significant for \sCh models above $\fehmath\gtrsim-1\,\dex$ which is driven by the neutron excess for more metal-rich models altering the relative abundance of iron-group elements from incomplete silicon burning (typically during He detonation). On the other hand, the $\xfemath{Mn}_\mathrm{Ia}$ for the \Ch DDT models \citep{Seitenzahl2013,Kobayashi2020} are much less metallicity dependent as in this case $^{55}$Co is typically produced in nuclear statistical equilibrium so is sensitive more to the neutron fraction from electron capture and less to the initial composition. Additionally, pre-explosion simmering \citep{PiroBildsten2008} should make the pre-explosion neutron abundance more insensitive to the initial metallicity. As highlighted by \cite{Gronow2021b}, the rising trend of \xfe{Mn} vs. \feh for the Milky Way then requires significant metallicity-dependent \sCh contributions. On the other hand, there are theoretical arguments that the \Ch channel becomes more significant for higher metallicity systems as winds are required to grow WDs stably to \Ch which are more effective at higher metallicity \citep{KobayashiNomoto2009}. In the recent Milky Way models of \citet[][Figure 14]{Kobayashi2020}, \Ch \ia only begin contributing after $t=4\,\mathrm{Gyr}$ when the Galaxy is enriched to around $\fehmath=-1\dex$. However, we must also consider the birth and evolutionary properties of binary systems with metallicity. Recently, it has become clear that binarity is more common at low metallicities \citep{Moe2019} possibly due to increased fragmentation at lower metallicity caused by lower dust opacities and enhanced gas cooling through dust coupling. Whether such an effect could differentially affect the two channels is not clear. Additionally, as highlighted by \cite{Ruiter2020}, mass-loss is more effective at higher metallicities. At very low metallicity ($\fehmath\approx-2\,\mathrm{dex}$) low mass-loss can produce fewer single-degenerate channel \ia as ONe WDs are formed preferentially. At higher metallicities, increased mass loss over the stellar lifetime can change the evolutionary pathway of a binary, make the Chandrasekhar mass harder to reach or prevent contact and subsequent stable mass transfer (as highlighted above). These three effects can significantly alter the predicted rates of different channels.

\subsubsection{Star-formation-history-dependent Type Ia yields}
\begin{figure*}
    \centering
    \includegraphics[width=0.8\linewidth]{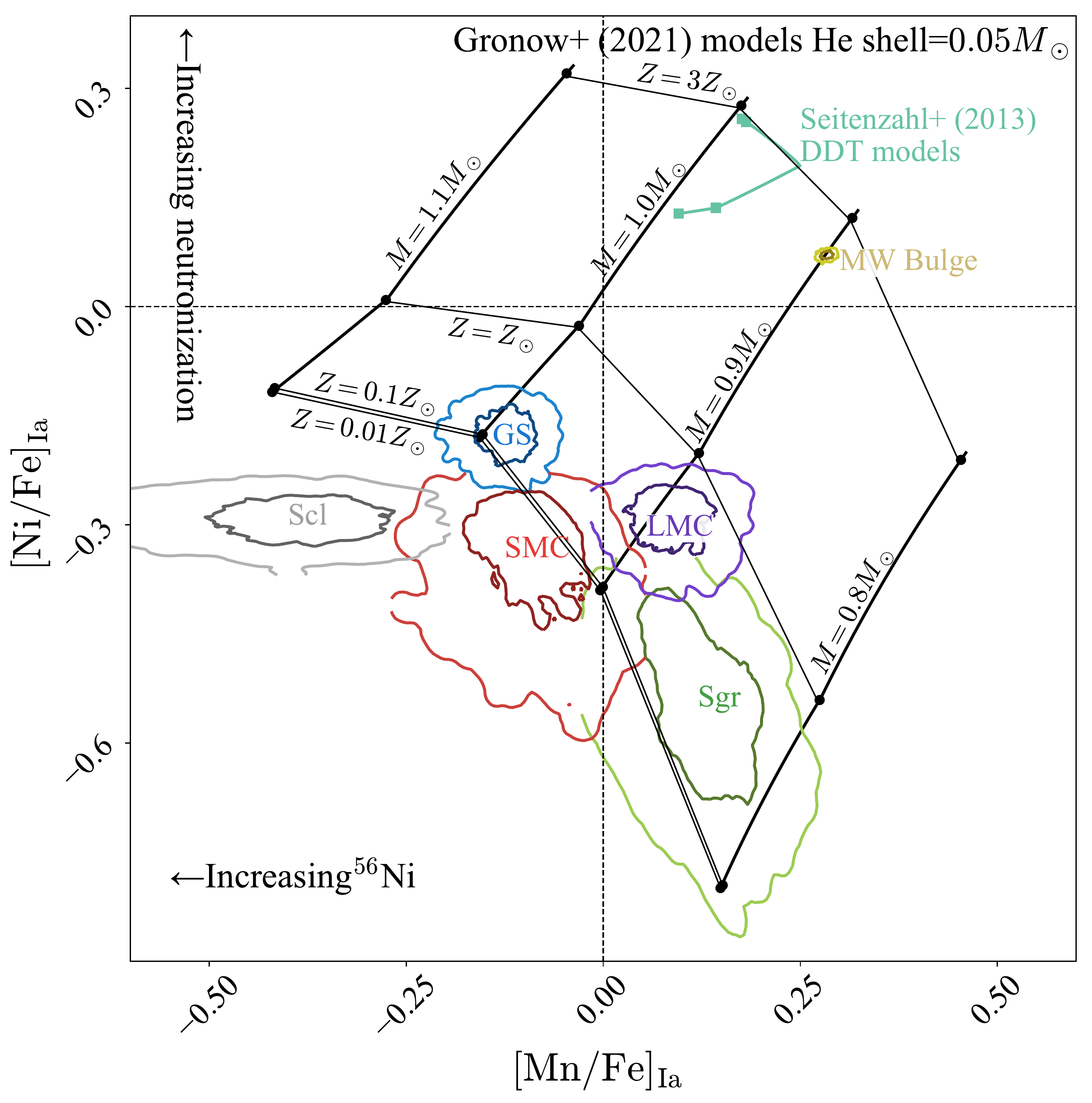}
    \caption{Joint constraints on Type Ia $\xfemath{Mn}$ and  $\xfemath{Ni}$ yields for a range of Local Group systems together with the \sCh models of \protect\citet[][black]{Gronow2021a,Gronow2021b} and the \Ch DDT models of \protect\citet[][green]{Seitenzahl2013}. The yields are functions of both core mass and metallicity of the \sCh progenitor. Increased mass increases $^{56}$Ni production (and hence Fe) whilst Mn remains constant (as it is formed in lower density regions at the core-shell interface), and increases the neutronization elevating the abundance of neutron-rich $^{58}$Ni.}
    \label{fig:mn_ni_sys}
\end{figure*}
Finally, we consider the possibility that some of the variation we see is due to different star formation histories in the systems. There is a well-studied trend for more massive galaxies to host intrinsically fainter \ia events \citep[e.g.][]{Howell2009}, which could be related to either age or metallicity effects. One plausible explanation is that the more massive systems host longer delay time \ia, which could arise from lower mass progenitor systems in the \sCh scenario \citep{Howell2011} although disentangling age and metallicity effects is non-trivial. This observation would appear at odds with the theory of increased dominance of \Ch systems in higher metallicity environments, although crucially the host galaxy correlations are against gas-phase metallicity which could be quite different from the progenitor metallicity depending on the star formation history. In the \sCh scenario there is the additional parameter of the progenitor mass. As noted previously, the models of \cite{Kobayashi2020} show weak mass dependence at fixed metallicity, whilst there is more significant variation with mass in the \cite{Gronow2021a,Gronow2021b} models. In Fig.~\ref{fig:met_dep} we show the $0.8,0.9$ and $1.1M_\odot$ core mass models from \cite{Gronow2021a,Gronow2021b} as thin lines. At fixed metallicity, the \xfe{Mn} yield increases with decreasing core mass. This is mostly driven by the correlation of $^{56}$Ni mass with core mass, as the mass of Mn is linked to the He detonation at the lower density core/shell interface and is more constant with core mass. We observe that systems such as the LMC or Sgr with perhaps more extended star formation histories may be entirely consistent with $100\,\percent$ \sCh models but of lower mass. Such a picture would be in line with the supernovae population studies. It appears from \xfe{Mn} alone that breaking the degeneracy between core mass and metallicity is not trivial. 

\cite{Flors2020} have used spectroscopic observations of the ratio $^{58}$Ni to $^{56}$Ni to infer the mass of individual \ia. As a neutron-rich species, the $^{58}$Ni abundance depends on neutron fraction so is higher for higher mass, higher density WDs \citep{Timmes2003}. This difference in behaviour with mass for \xfe{Ni} compared to \xfe{Mn} allows breaking of the degeneracy when using only \xfe{Mn}. In Fig.~\ref{fig:mn_ni_sys} we display the joint constraints on Type Ia \xfe{Mn} and \xfe{Ni} yields for the six best constrained systems we have analysed alongside the \sCh models from \citet{Gronow2021a,Gronow2021b}. We observe that, despite higher Type Ia \xfe{Mn} abundance in the LMC and Sgr compared to the \gs, the \xfe{Ni} abundance in these systems is lower than that for the \gs and appears more consistent with $0.8$ or $0.9M_\odot$ core mass models, in line with our expectation from their more extended star formation history. We have chosen the $0.05M_\odot$ He shell mass models from  \citet{Gronow2021a,Gronow2021b} but similar conclusions hold for other He shell mass models although there is more variance in the abundances for lower ($0.8,0.9M_\odot$) core mass models with varying shell mass. To draw firmer conclusions requires a more detailed modelling of the star formation histories of the systems and the delay-time distributions with WD mass, but the combined constraints from \xfe{Mn} and \xfe{Ni} appear to provide a route to more strongly constrain the \ia progenitor scenario and could even point towards \sCh dominance in all systems.

Based on the currently available theoretical models, the combination of results from the different systems suggest an increasing significance of \Ch explosions with increasing metallicity, although the differences between different theoretical models hint there may be a model where the metallicity dependence is explained entirely by \sCh explosions, possibly with a distribution of masses related to the star formation history of the system. The plane of \xfe{Mn} vs. \xfe{Ni} seems ideal to distinguish these scenarios.

\section{Conclusions}\label{sec::conclusions}
The range of different \ia scenarios are distinguished by different nucleosynthetic yields, which are in turn imprinted in the chemical abundances of stars. \ia channels potentially vary with the mass, metallicity, initial-mass-function and star-formation-history of the host systems which could have profound consequences on the use of \ia as cosmological probes. However, high-quality resolved stellar spectroscopic abundances are available only for the Milky Way and several Local Group dwarf galaxies. Recently, it has been discovered that a substantial fraction of the Milky Way's halo consists of a population of kinematically and chemically distinct stars consistent with originating from a significant merger, dubbed the \gs, early in the life of the Milky Way. This gives us a unique probe into a low metallicity, early-Universe galaxy using sets of bright local tracers.

We have taken three samples of \gs stars from \cite{NS10}, GALAH DR3 \citep{Buder2020} and APOGEE DR16 \citep{SDSSDR16} for which high-dimensional chemical data are available (8+ abundances), and fitted their chemical-space distributions with the simple analytic chemical evolution models of \cite{Weinberg2017} to measure the abundance signatures of \ia. We find the down-turning $\xfemath{\alpha}$ sequence indicative of the onset of \ia is accompanied by a relatively flat \xfe{Mn} and weakly declining \xfe{Ni} trend indicating low production in \ia ($\sim-0.15\dex$ and $\sim-0.3\dex$ respectively). Comparison with theoretical nucleosynthetic yields for \ia exploded as \Ch pure deflagrations, \Ch deflagration-to-detonation transition and double-detonation \sCh demonstrate such behaviour is only possible with a significant ($\gtrsim60\,\percent$) contribution of \sCh systems, with some theoretical calculations implying near $100\,\percent$ \sCh systems. This corroborates findings from dwarf galaxies of the Local Group. 

Finally, we concluded our study with a combined analysis of several systems with varying star formation histories and metallicity distributions (Milky Way bulge, Magellanic clouds, \gs, dwarf spheroidal galaxies). We found to reproduce the chemical enrichment of all of these systems requires a metallicity-dependent Type Ia \xfe{Mn} yield. Only \sCh models have significant metallicity-dependent yields. This then indicates either \begin{inparaenum}
\item a significant contribution from \sCh \ia in all systems, possibly with variation in the WD mass distribution with star formation history which could be elucidated using a combination of \xfe{Ni} and \xfe{Mn} measurements, or \item an enhancement of the \Ch channel at higher metallicity, possibly due to stabilising winds allowing white dwarf growth to the Chandrasekhar mass.
\end{inparaenum}

We close by reflecting on the implications of our work on using \ia as cosmological probes. The understanding of the progenitors of \ia is an important part of using these objects as standard candles. However, the success in their use has arisen from the ability to standardise the variety of observed light curves, even despite an understanding of what drives the variation. From the \Ch perspective, the free parameter is perhaps the time a deflagration flame transitions into a detonation wave \citep{Blondin2013}, whilst in the \sCh picture, it is the mass of the WD. In both cases, however, the primary driver in light curve variation is then the $^{56}$Ni production which gives rise to the width-luminosity relation \citep{Scalzo2014}. Beyond this, there is an intrinsic scatter of $\sim0.15\,\mathrm{mag}$ in \ia luminosities, the cause of which is not understood. If the residual depends on the progenitor scenario, and possibly more crucially the host galaxy properties, there is scope to improve the calibrations and remove potential systematics \citep[e.g.][]{Sullivan2010}. For instance, several studies \citep[e.g.][]{Howell2009,Kelly2010,Scalzo2014} have demonstrated \ia in less massive galaxies tend to be brighter (even after calibrating the light curves), reflecting possible age \citep{Childress2014} or metallicity effects \citep{Timmes2003} on the progenitors. Such correlations favour the \sCh channel where lower mass older WDs could explode at longer delay times producing lower luminosity SNe Ia \citep{Howell2011}. Such studies are concerned with the properties of observed supernovae, whilst here we are only sensitive to the \ia which enrich future star-forming gas, and furthermore, supernovae observations are unable to probe the gas metallicity when the progenitor was born. For these reasons, our results are complementary to \ia population studies and taken together give a route into further understanding the \ia progenitor properties with both metallicity and star formation history of the host. 

\section*{Acknowledgements}
The authors would like to thank Chiaki Kobayashi for providing the yields from the work of \cite{Kobayashi2020}. J.L.S. acknowledges support from the Royal Society (URF\textbackslash R1\textbackslash191555). Thanks to Richard Ellis for suggesting investigating supernovae scenarios using \gs data.

This work made use of the Heidelberg Supernova Model Archive (HESMA), https://hesma.h-its.org.
This paper made use of
\textsc{numpy} \citep{numpy},
\textsc{scipy} \citep{scipy}, 
\textsc{matplotlib} \citep{matplotlib}, 
\textsc{seaborn} \citep{seaborn},
\textsc{astropy} \citep{astropy:2013,astropy:2018}, \textsc{ChainConsumer} \citep{chainconsumer} and
\textsc{Stan} \citep{stan}.
This work has made use of data from the European Space Agency (ESA) mission
{\it Gaia} (\url{https://www.cosmos.esa.int/gaia}), processed by the {\it Gaia}
Data Processing and Analysis Consortium (DPAC,
\url{https://www.cosmos.esa.int/web/gaia/dpac/consortium}). Funding for the DPAC
has been provided by national institutions, in particular the institutions
participating in the {\it Gaia} Multilateral Agreement.

This work made use of the Third Data Release of the GALAH Survey \citep{Buder2020}. The GALAH Survey is based on data acquired through the Australian Astronomical Observatory, under programs: A/2013B/13 (The GALAH pilot survey); A/2014A/25, A/2015A/19, A2017A/18 (The GALAH survey phase 1); A2018A/18 (Open clusters with HERMES); A2019A/1 (Hierarchical star formation in Ori OB1); A2019A/15 (The GALAH survey phase 2); A/2015B/19, A/2016A/22, A/2016B/10, A/2017B/16, A/2018B/15 (The HERMES-TESS program); and A/2015A/3, A/2015B/1, A/2015B/19, A/2016A/22, A/2016B/12, A/2017A/14 (The HERMES K2-follow-up program). We acknowledge the traditional owners of the land on which the AAT stands, the Gamilaraay people, and pay our respects to elders past and present. This paper includes data that has been provided by AAO Data Central (\url{datacentral.aao.gov.au}).

Funding for the Sloan Digital Sky 
Survey IV has been provided by the 
Alfred P. Sloan Foundation, the U.S. 
Department of Energy Office of 
Science, and the Participating 
Institutions. 
SDSS-IV acknowledges support and 
resources from the Center for High 
Performance Computing  at the 
University of Utah. The SDSS 
website is \url{www.sdss.org}.
SDSS-IV is managed by the 
Astrophysical Research Consortium 
for the Participating Institutions 
of the SDSS Collaboration including 
the Brazilian Participation Group, 
the Carnegie Institution for Science, 
Carnegie Mellon University, Center for 
Astrophysics | Harvard \& 
Smithsonian, the Chilean Participation 
Group, the French Participation Group, 
Instituto de Astrof\'isica de 
Canarias, The Johns Hopkins 
University, Kavli Institute for the 
Physics and Mathematics of the 
Universe (IPMU) / University of 
Tokyo, the Korean Participation Group, 
Lawrence Berkeley National Laboratory, 
Leibniz Institut f\"ur Astrophysik 
Potsdam (AIP),  Max-Planck-Institut 
f\"ur Astronomie (MPIA Heidelberg), 
Max-Planck-Institut f\"ur 
Astrophysik (MPA Garching), 
Max-Planck-Institut f\"ur 
Extraterrestrische Physik (MPE), 
National Astronomical Observatories of 
China, New Mexico State University, 
New York University, University of 
Notre Dame, Observat\'ario 
Nacional / MCTI, The Ohio State 
University, Pennsylvania State 
University, Shanghai 
Astronomical Observatory, United 
Kingdom Participation Group, 
Universidad Nacional Aut\'onoma 
de M\'exico, University of Arizona, 
University of Colorado Boulder, 
University of Oxford, University of 
Portsmouth, University of Utah, 
University of Virginia, University 
of Washington, University of 
Wisconsin, Vanderbilt University, 
and Yale University.
\section*{Data Availability}
All of the data used in this work is in the public domain.


\bibliographystyle{mnras}
\bibliography{bibliography} 



\appendix
\section{Common stars}\label{appendix::common_stars}
\begin{table*}
    \caption{Offset parameters between common stars observed by \protect\citetalias{SDSSDR16} and \protect\citetalias{Buder2020}. Models of the difference between APOGEE and GALAH (i.e. positive numbers mean APOGEE is larger than GALAH) have been fitted with the linear models of the form $c_i+m_{\teff}(\teff/1000\,\mathrm{K}-4.5)+m_{\logg}(\logg-2.5)+m_{\fehmath}(\fehmath+1)$. The top set of fits correspond to a constant model $c_0$ ($m_{\teff}=m_{\logg}=m_{\fehmath}=0$) and three models where the gradient in only one parameter is fitted for (e.g. $c_{\teff},m_{\teff}$ correspond to the case where ($m_{\logg}=m_{\fehmath}=0$). The second section shows the results for a full multivariate fit. The final section shows the results for fitting a metallicity-dependent model to the Ni differences using the three different line combinations. The spectroscopic parameters $(\teff, \logg, \fehmath)$ used are those of GALAH DR3 although they differ from APOGEE DR16 parameters by $3.6\,\mathrm{K}, -0.002\,\mathrm{dex}$ and $-0.05\,\mathrm{dex}$ in the median.}
    \centering
    \begin{tabular}{l|c|cc|cc|cc}
&$c_0$&$c_{\teff}$&$m_{\teff}$
&$c_{\logg}$&$m_{\logg}$
&$c_{\fehmath}$&$m_{\fehmath}$\\
\hline
Mg&$+0.04^{+0.01}_{-0.01}$&$+0.03^{+0.01}_{-0.01}$&$+0.04^{+0.02}_{-0.02}$&$+0.04^{+0.01}_{-0.01}$&$-0.00^{+0.01}_{-0.01}$&$+0.04^{+0.01}_{-0.01}$&$-0.16^{+0.02}_{-0.02}$\\
Si&$-0.01^{+0.01}_{-0.01}$&$-0.01^{+0.01}_{-0.01}$&$-0.01^{+0.02}_{-0.02}$&$-0.02^{+0.01}_{-0.01}$&$-0.04^{+0.01}_{-0.01}$&$-0.01^{+0.01}_{-0.01}$&$-0.11^{+0.03}_{-0.03}$\\
Ca&$-0.05^{+0.01}_{-0.01}$&$-0.06^{+0.01}_{-0.01}$&$+0.04^{+0.02}_{-0.02}$&$-0.04^{+0.01}_{-0.01}$&$+0.03^{+0.01}_{-0.01}$&$-0.06^{+0.01}_{-0.01}$&$+0.01^{+0.03}_{-0.03}$\\
Cr&$+0.02^{+0.01}_{-0.01}$&$+0.02^{+0.01}_{-0.01}$&$+0.04^{+0.03}_{-0.03}$&$+0.03^{+0.02}_{-0.02}$&$+0.01^{+0.02}_{-0.02}$&$+0.04^{+0.01}_{-0.01}$&$-0.18^{+0.05}_{-0.05}$\\
Mn&$+0.03^{+0.01}_{-0.01}$&$+0.05^{+0.01}_{-0.01}$&$-0.12^{+0.02}_{-0.02}$&$-0.00^{+0.01}_{-0.01}$&$-0.07^{+0.01}_{-0.01}$&$+0.02^{+0.01}_{-0.01}$&$+0.03^{+0.05}_{-0.05}$\\
Ni&$+0.06^{+0.01}_{-0.01}$&$+0.04^{+0.01}_{-0.01}$&$+0.11^{+0.02}_{-0.02}$&$+0.07^{+0.01}_{-0.01}$&$+0.02^{+0.01}_{-0.01}$&$+0.06^{+0.01}_{-0.01}$&$-0.13^{+0.03}_{-0.03}$\\
Cu&$+0.35^{+0.03}_{-0.03}$&$+0.40^{+0.03}_{-0.03}$&$-0.29^{+0.07}_{-0.08}$&$+0.26^{+0.02}_{-0.03}$&$-0.23^{+0.03}_{-0.03}$&$+0.46^{+0.02}_{-0.02}$&$-1.20^{+0.12}_{-0.11}$\\
\hline
Mg&$-0.02^{+0.02}_{-0.02}$&-&$+0.15^{+0.02}_{-0.02}$&-&$-0.07^{+0.02}_{-0.02}$&-&$-0.12^{+0.04}_{-0.03}$\\
Si&$-0.14^{+0.02}_{-0.02}$&-&$+0.31^{+0.02}_{-0.03}$&-&$-0.17^{+0.02}_{-0.02}$&-&$+0.01^{+0.05}_{-0.04}$\\
Ca&$-0.02^{+0.01}_{-0.01}$&-&$-0.05^{+0.03}_{-0.03}$&-&$+0.05^{+0.01}_{-0.01}$&-&$-0.04^{+0.03}_{-0.03}$\\
Cr&$+0.05^{+0.03}_{-0.03}$&-&$-0.03^{+0.06}_{-0.06}$&-&$+0.02^{+0.04}_{-0.04}$&-&$-0.19^{+0.08}_{-0.07}$\\
Mn&$-0.02^{+0.02}_{-0.02}$&-&$+0.01^{+0.05}_{-0.04}$&-&$-0.07^{+0.02}_{-0.02}$&-&$+0.08^{+0.05}_{-0.05}$\\
Ni&$+0.05^{+0.01}_{-0.01}$&-&$+0.14^{+0.03}_{-0.03}$&-&$-0.01^{+0.02}_{-0.02}$&-&$-0.14^{+0.04}_{-0.04}$\\
Cu&$+0.21^{+0.06}_{-0.06}$&-&$+0.62^{+0.11}_{-0.12}$&-&$-0.35^{+0.07}_{-0.06}$&-&$-0.97^{+0.13}_{-0.14}$\\
\hline
Ni 5847 \AA&-&-&-&-&-&$+0.02^{+0.09}_{-0.08}$&$+0.29^{+0.36}_{-0.41}$\\
Ni 6586 \AA&-&-&-&-&-&$+0.09^{+0.02}_{-0.02}$&$-0.01^{+0.05}_{-0.05}$\\
Ni Both&-&-&-&-&-&$+0.06^{+0.01}_{-0.01}$&$-0.17^{+0.03}_{-0.03}$\\
    \end{tabular}
    \label{tab:models_offsets}
\end{table*}

\begin{figure*}
    \centering
    \includegraphics[width=\textwidth]{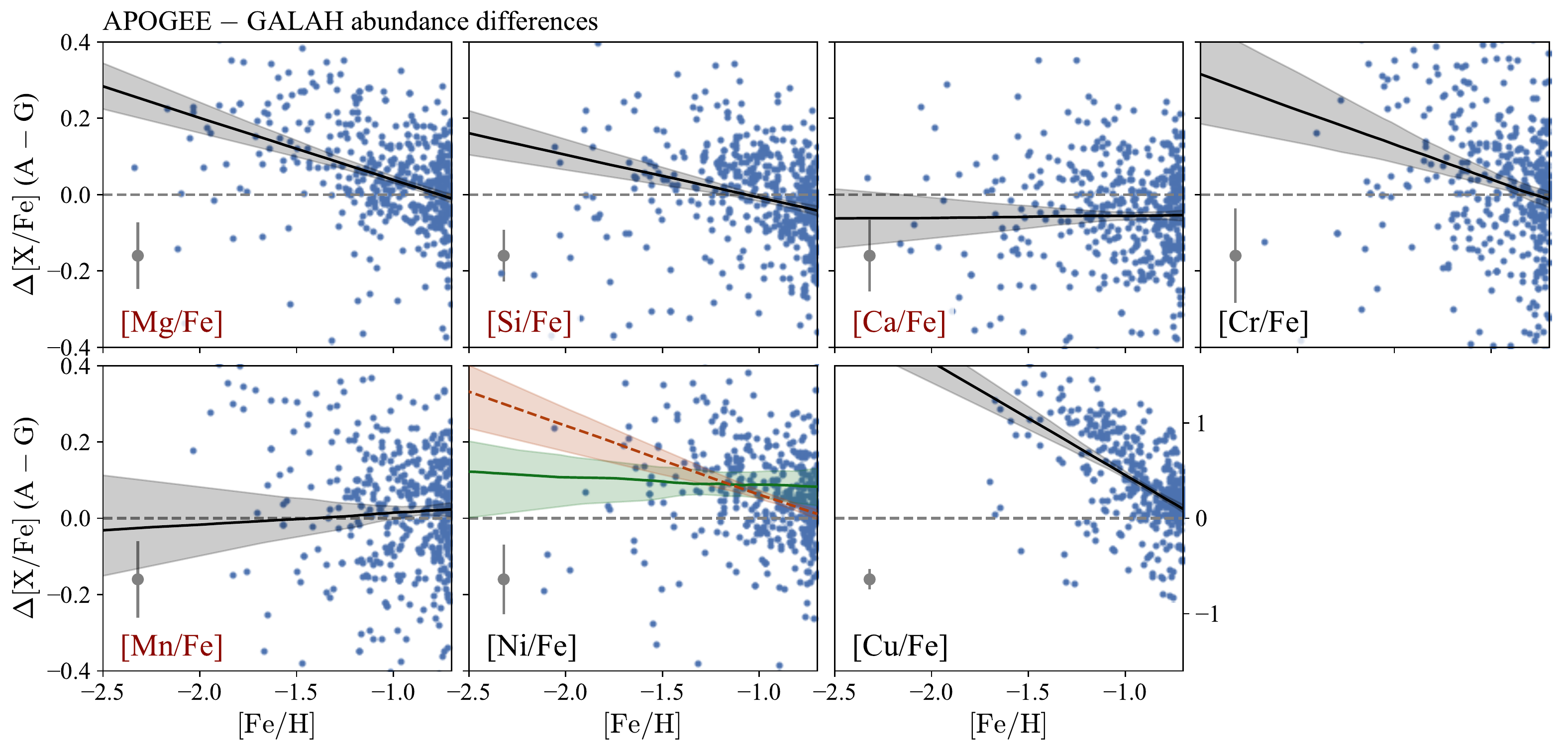}
    \caption{Differences between the \protect\citetalias{SDSSDR16} and \protect\citetalias{Buder2020} abundances plotted against \feh. Simple linear models are plotted (mostly in black with a $1\sigma$ envelope). The parameters of these models are in Table~\ref{tab:models_offsets}. For \xfe{Ni} we show two fits for the GALAH abundances derived from Ni $6586$\AA\ in solid green and for those derived from both Ni $5847$\AA\ and Ni $6586$\AA\ in dashed red. Median errors for the sample are given in the lower left.}
    \label{fig:a_vs_g}
\end{figure*}

\begin{figure*}
    \centering
    \includegraphics[width=\columnwidth]{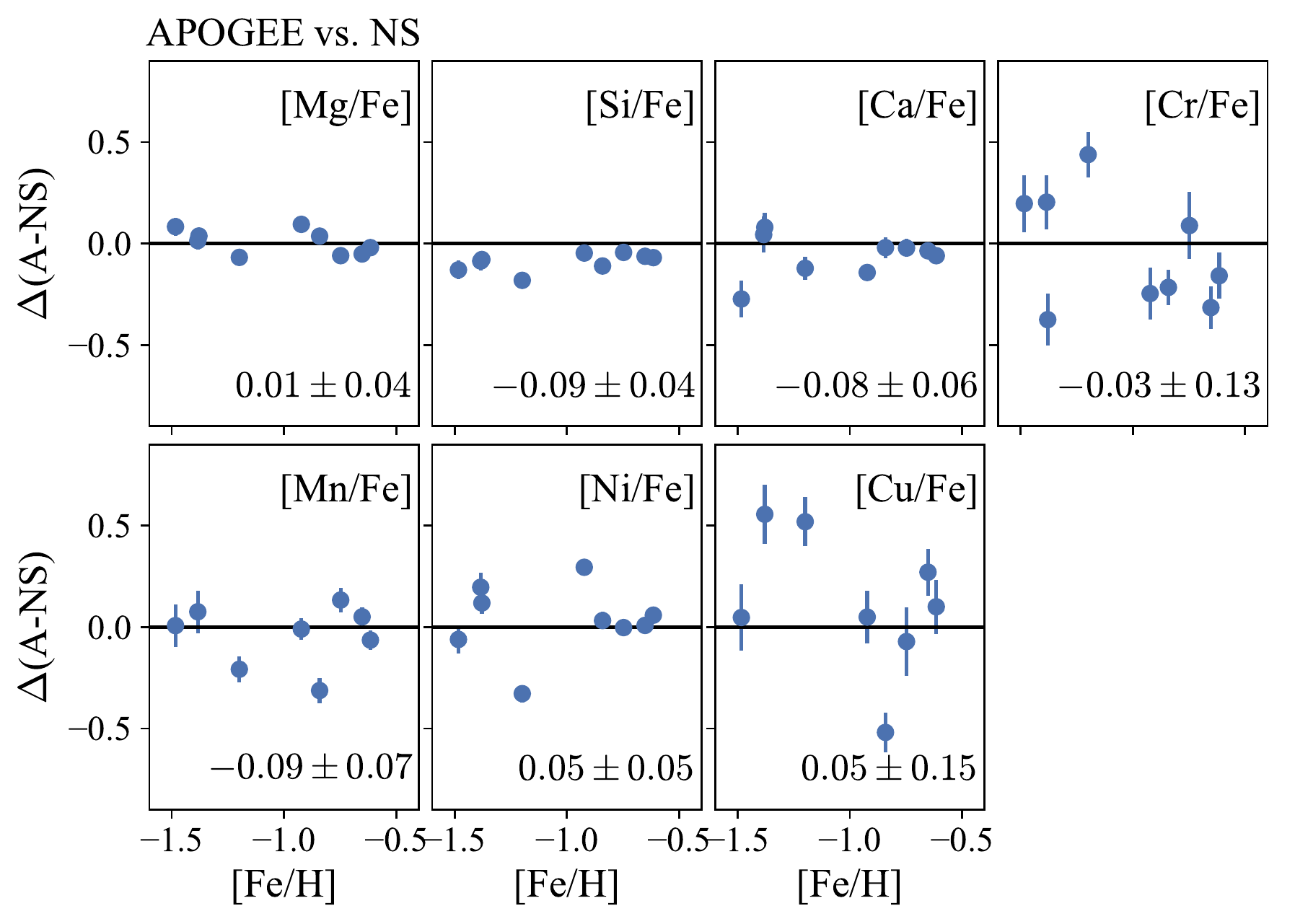}
    \includegraphics[width=\columnwidth]{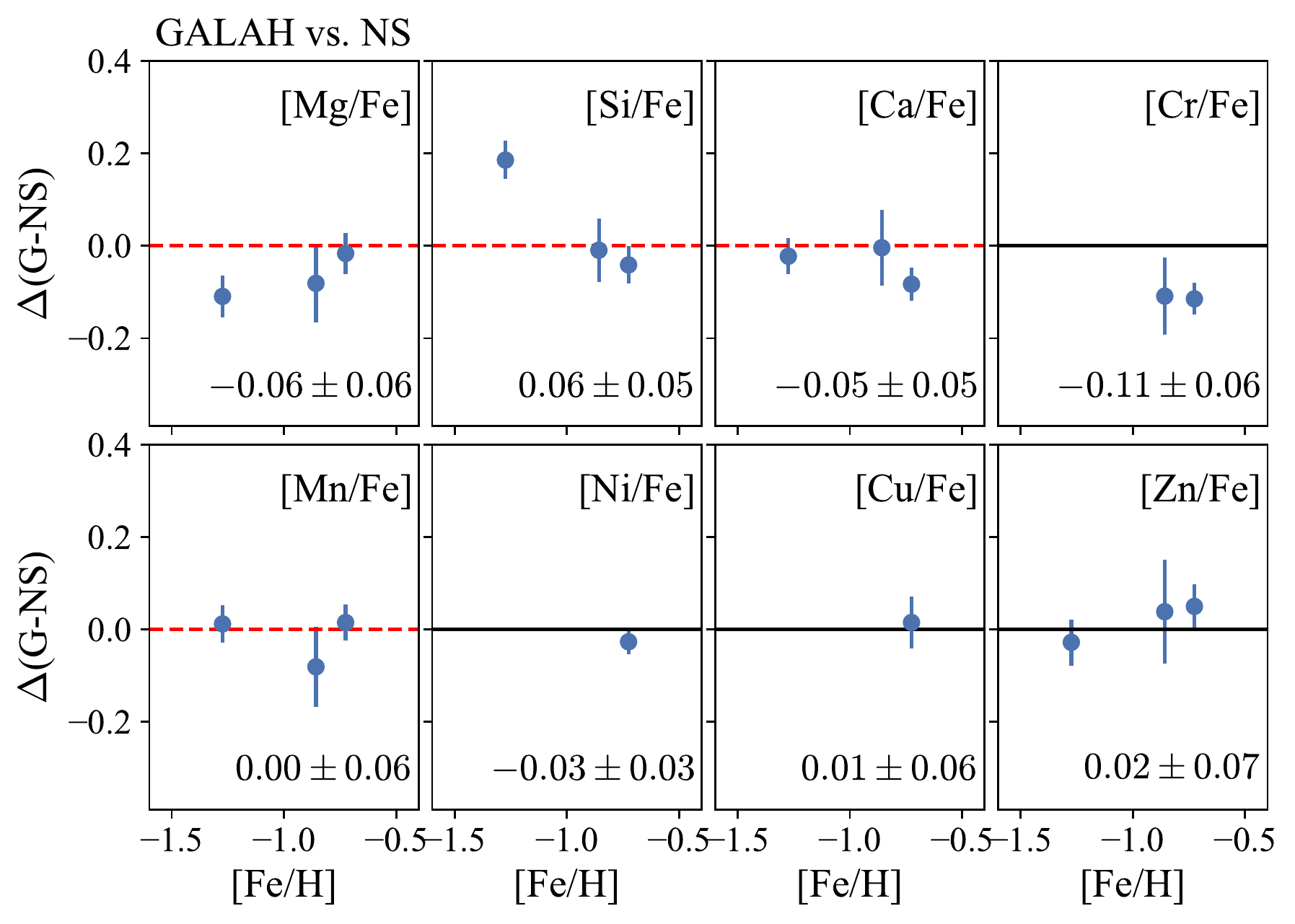}
    \caption{\protect\citetalias{SDSSDR16} (A, left) and \protect\citetalias{Buder2020} (G, right) vs. \protect\citetalias{NS10}  abundances: we display the difference in abundance measurements for stars in common. The inverse-variance-weighted mean with uncertainty is displayed in each panel. Panels with red dashed horizontal lines have 1D non-LTE corrections applied in GALAH. Note the different scales for the left and right sets of panels.}
    \label{fig:ns_vs_ag}
\end{figure*}

For the analysis in this paper, it is important that the absolute abundance scale of each dataset is well determined, and that the stellar samples are free from systematic trends with stellar parameters (in particular, metallicity). As we have access to multiple datasets we are able to compare abundances for stars observed by two surveys. We initially compare the \citetalias{SDSSDR16} and \citetalias{Buder2020} datasets as they have the largest overlap, and then we go on to compare these datasets with \citetalias{NS10}.

\subsection{APOGEE vs. GALAH}
We take the same sample of \citetalias{Buder2020} as detailed in Section~\ref{sec::data_galah} without restricting to just \gs stars, and cross-match to \citetalias{SDSSDR16} stars with no ASPCAP flags. This results in $479$ stars. We fit a series of linear models to the difference in the Mg, Si, Ca, Cr, Mn, Ni and Cu abundances (only using unflagged abundances) as a function of the stellar parameters ($T_\mathrm{eff}$, $\log g$ and $\fehmath$). Note that APOGEE Cu is known to be unreliable particularly at low metallicity \citep{Jonsson2020} so we don't use it in the main body of the paper. We account for the uncertainties in the abundances but not in the stellar parameters (which on the whole are small). Importantly we account for the floor in the APOGEE \xh{X} abundances below which abundances aren't reported. Excluding this effect biases the trends. 

In Table~\ref{tab:models_offsets} we show the results of fitting linear models to the abundance differences as a function of $T_\mathrm{eff}$, $\log g$ and $\fehmath$ separately, and using all three together. The results of the fits against $\fehmath$ are shown in Fig.~\ref{fig:a_vs_g}. The results at the metal-rich end are mostly consistent with $\lesssim0.1\dex$ difference. \cite{Jonsson2020} show how the majority of APOGEE abundances agree (within $\sim0.05\dex$) in the median with those obtained from optical spectra (using LTE). For Mg and Si we find significant trends with metallicity with APOGEE abundances larger than GALAH at the low metallicity end. Both of these elements have non-LTE corrections in GALAH \citep{Amarsi2020}. Ca also has non-LTE corrections in GALAH but we don't observe a trend, just an offset of $-0.06\dex$. \cite{Osorio2020} found that non-LTE corrections for Ca and Mg in the Sun, Arcturus and Procyon are smaller in the H-band than in the optical, and always $\lesssim0.1\dex$. For Cr and Cu, we find a significant trend with APOGEE overestimating relative to GALAH at the metal-poor end. However, Cu in APOGEE is known to be unreliable \citep{Jonsson2020}. For Mn we find essentially no bias between APOGEE and GALAH across the full metallicity range. This is despite the non-LTE corrections in GALAH. This gives confidence to the conclusions in the main body of the paper which use the Mn abundances. Finally, with Ni the GALAH analysis used different combinations of the $5847$\AA\ and $6586$\AA\ lines. We found different trends with respect to the APOGEE Ni abundances when restricting to subsets of GALAH abundances derived from the same combinations. In particular, we found a significant trend when using only $5847$\AA\ whilst no trend for only $6586$\AA. This suggests there is a systematic variation in the GALAH Ni abundances.

\subsection{APOGEE and GALAH vs. Nissen \& Schuster}
We find there are three stars in the \citetalias{NS10} sample also observed as part of \citetalias{Buder2020}, and nine also observed in \citetalias{SDSSDR16}. In Fig.~\ref{fig:ns_vs_ag} we plot the difference in the reported abundances and report the inverse-variance-weighted mean difference. Despite the small number of datapoints, we find good correspondence between the two datasets with only \xfe{Cr} showing a significant offset (\citetalias{Buder2020} smaller than \citetalias{NS10}). Six of the inspected abundances have 1D non-LTE corrections in \citetalias{Buder2020} \citep{Amarsi2020} whilst \citetalias{NS10} use an LTE analysis \citep[except in the case of \xfe{Cu} where we have used the results from][]{Yan2016}.

\section{Auxiliary data tables}
\begin{figure*}
    \centering
    \includegraphics[width=0.92\textwidth]{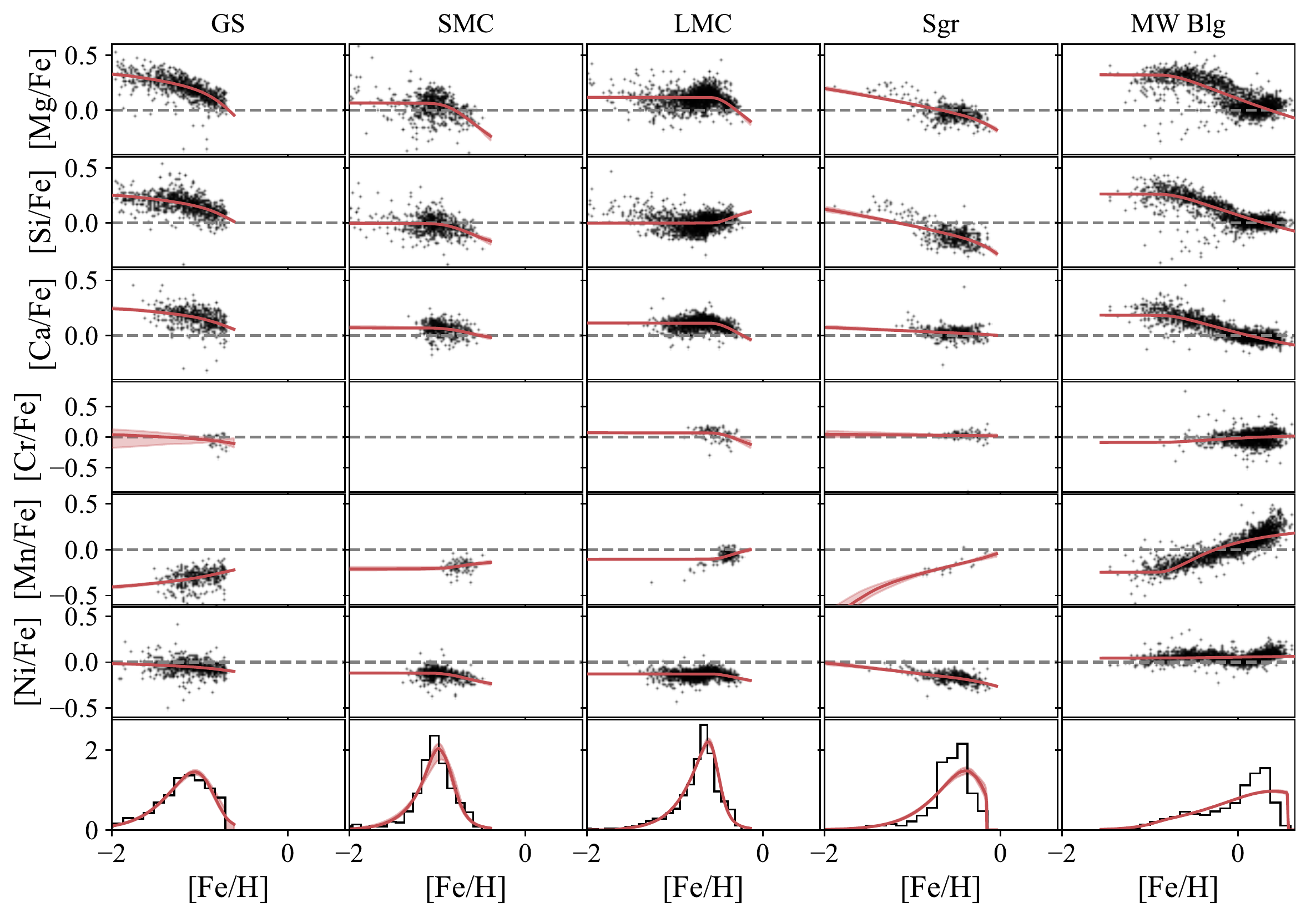}
    \caption{Chemical evolution model fits for \gs, SMC, LMC, Sgr and the Milky Way bulge. The data for each abundance \xfe{X} vs. metallicity are shown in black with the bottom row showing the metallicity distribution of the data. The red lines show the median model fits and the bands the $16,84$th percentile confidence interval.}
    \label{fig:lg_model_comparison}
\end{figure*}

Here we provide additional data tables from out study. The results using the exponential star formation rate of \cite{Weinberg2017} applied to the \citetalias{NS10} data are shown in Table~\ref{tab::model_difference}. The full set of constrained parameters for the Local Group systems are shown in Table~\ref{tab::lg_fullparams}. We show the corresponding models for the most massive five systems (\gs, Sgr, SMC, LMC and the Milky Way bulge) in Fig.~\ref{fig:lg_model_comparison}.

\begin{table}
    \centering
    \caption{Comparison between derived yields using \protect\cite{NS10} dataset with linear-exponential star formation model (NS10) and the exponential model (NS10 Exp Model) as detailed by \protect\cite{Weinberg2017}.}
    \begin{tabular}{llll}
Type&Element&NS10&NS10 Exp Model\\\hline
Ia&$\log_{10}m_\mathrm{Fe}$&$-2.73^{+0.10}_{-0.10}$&$-2.77^{+0.10}_{-0.10}$\\
Ia&$[\mathrm{Mg}/\mathrm{Fe}]$&$-2.00^{+0.35}_{-0.34}$&$-1.99^{+0.33}_{-0.34}$\\
Ia&$[\mathrm{Si}/\mathrm{Fe}]$&$-0.87^{+0.34}_{-0.54}$&$-0.85^{+0.34}_{-0.52}$\\
Ia&$[\mathrm{Ca}/\mathrm{Fe}]$&$-0.72^{+0.31}_{-0.52}$&$-0.68^{+0.29}_{-0.50}$\\
Ia&$[\mathrm{Cr}/\mathrm{Fe}]$&$+0.03^{+0.03}_{-0.03}$&$+0.02^{+0.03}_{-0.03}$\\
Ia&$[\mathrm{Mn}/\mathrm{Fe}]$&$-0.16^{+0.03}_{-0.03}$&$-0.17^{+0.03}_{-0.03}$\\
Ia&$[\mathrm{Ni}/\mathrm{Fe}]$&$-0.41^{+0.08}_{-0.12}$&$-0.41^{+0.08}_{-0.11}$\\
Ia&$[\mathrm{Cu}/\mathrm{Fe}]$&$-0.46^{+0.19}_{-0.35}$&$-0.49^{+0.21}_{-0.35}$\\
Ia&$[\mathrm{Zn}/\mathrm{Fe}]$&$-0.91^{+0.31}_{-0.52}$&$-0.90^{+0.31}_{-0.49}$\\
\hline
II&$\log_{10}m_\mathrm{Fe}$&$-3.08^{+0.16}_{-0.16}$&$-2.93^{+0.14}_{-0.15}$\\
II&$[\mathrm{Mg}/\mathrm{Fe}]$&$+0.48^{+0.11}_{-0.09}$&$+0.41^{+0.09}_{-0.07}$\\
II&$[\mathrm{Si}/\mathrm{Fe}]$&$+0.46^{+0.10}_{-0.09}$&$+0.40^{+0.09}_{-0.06}$\\
II&$[\mathrm{Ca}/\mathrm{Fe}]$&$+0.54^{+0.10}_{-0.08}$&$+0.47^{+0.09}_{-0.06}$\\
II&$[\mathrm{Cr}/\mathrm{Fe}]$&$-0.07^{+0.04}_{-0.06}$&$-0.05^{+0.02}_{-0.04}$\\
II&$[\mathrm{Mn}/\mathrm{Fe}]$&$-0.56^{+0.11}_{-0.27}$&$-0.47^{+0.06}_{-0.14}$\\
II&$[\mathrm{Ni}/\mathrm{Fe}]$&$+0.08^{+0.08}_{-0.06}$&$+0.03^{+0.07}_{-0.04}$\\
II&$[\mathrm{Cu}/\mathrm{Fe}]$&$-0.41^{+0.19}_{-0.35}$&$-0.39^{+0.14}_{-0.23}$\\
II&$[\mathrm{Zn}/\mathrm{Fe}]$&$+0.29^{+0.10}_{-0.08}$&$+0.22^{+0.09}_{-0.06}$\\
    \end{tabular}
    \label{tab::model_difference}
\end{table}

\setlength{\tabcolsep}{3pt}
\begin{table*}
\caption{Type Ia and Type II supernovae abundances and evolutionary parameters for chemical evolution models fitted to a range of Local Group systems. We give the median metallicity of the stars used to model each system.}
\begin{tabular}{llllllllllll}
Type&Element&UMi&CVnI&Leo I&Scl&GS&For&SMC&LMC&Sgr&MW Blg\\\hline
&[Fe/H]&-2.21&-1.98&-1.43&-1.63&-1.17&-1.07&-1.04&-0.69&-0.53&0.08\\\hline
Ia&$\log_{10}m_\mathrm{Fe}$&$-2.91^{+0.08}_{-0.08}$&$-2.80^{+0.09}_{-0.09}$&$-2.76^{+0.09}_{-0.10}$&$-2.65^{+0.09}_{-0.09}$&$-2.85^{+0.09}_{-0.09}$&$-2.89^{+0.09}_{-0.10}$&$-2.93^{+0.09}_{-0.09}$&$-2.99^{+0.10}_{-0.10}$&$-2.78^{+0.11}_{-0.10}$&$-2.30^{+0.05}_{-0.08}$\\
Ia&$[\mathrm{Mg}/\mathrm{Fe}]$&$-2.01^{+0.33}_{-0.35}$&$-2.00^{+0.33}_{-0.34}$&$-2.02^{+0.33}_{-0.34}$&$-2.02^{+0.34}_{-0.33}$&$-1.97^{+0.37}_{-0.36}$&$-1.98^{+0.37}_{-0.35}$&$-0.60^{+0.08}_{-1.29}$&$-0.58^{+0.06}_{-0.64}$&$-1.29^{+0.72}_{-0.72}$&$-0.507^{+0.006}_{-0.013}$\\
Ia&$[\mathrm{Si}/\mathrm{Fe}]$&$-0.22^{+0.18}_{-0.29}$&$-1.04^{+0.45}_{-0.56}$&$-1.05^{+0.34}_{-0.48}$&$-1.36^{+0.35}_{-0.48}$&$-0.32^{+0.05}_{-0.06}$&$-0.30^{+0.17}_{-0.25}$&$-0.29^{+0.06}_{-0.13}$&$+0.19^{+0.02}_{-0.02}$&$-1.44^{+0.49}_{-0.49}$&$-0.38^{+0.02}_{-0.02}$\\
Ia&$[\mathrm{Ca}/\mathrm{Fe}]$&$-0.31^{+0.18}_{-0.36}$&$-0.36^{+0.21}_{-0.44}$&$-0.19^{+0.05}_{-0.07}$&$-0.23^{+0.05}_{-0.05}$&$-0.17^{+0.05}_{-0.06}$&$-0.55^{+0.29}_{-0.48}$&$-0.07^{+0.03}_{-0.06}$&$-0.28^{+0.05}_{-0.08}$&$-0.05^{+0.02}_{-0.03}$&$-0.29^{+0.01}_{-0.01}$\\
Ia&$[\mathrm{Cr}/\mathrm{Fe}]$&$-1.04^{+0.39}_{-0.56}$&$-0.01^{+0.15}_{-0.42}$&$+0.63^{+0.05}_{-0.04}$&$+0.00^{+0.03}_{-0.03}$&$-0.23^{+0.31}_{-0.55}$&$+0.65^{+0.10}_{-0.09}$&$-0.01^{+0.83}_{-0.83}$&$-0.53^{+0.28}_{-0.49}$&$+0.00^{+0.07}_{-0.07}$&$+0.054^{+0.009}_{-0.009}$\\
Ia&$[\mathrm{Mn}/\mathrm{Fe}]$&$-0.44^{+0.22}_{-0.39}$&$-0.29^{+0.19}_{-0.36}$&$-0.20^{+0.11}_{-0.14}$&$-0.38^{+0.06}_{-0.07}$&$-0.12^{+0.02}_{-0.02}$&$-0.17^{+0.24}_{-0.42}$&$-0.09^{+0.04}_{-0.04}$&$+0.09^{+0.03}_{-0.03}$&$+0.13^{+0.05}_{-0.05}$&$+0.284^{+0.005}_{-0.005}$\\
Ia&$[\mathrm{Ni}/\mathrm{Fe}]$&$-0.91^{+0.44}_{-0.53}$&$-0.47^{+0.12}_{-0.28}$&$+0.11^{+0.03}_{-0.03}$&$-0.29^{+0.02}_{-0.02}$&$-0.18^{+0.02}_{-0.02}$&$+0.36^{+0.07}_{-0.07}$&$-0.32^{+0.03}_{-0.08}$&$-0.29^{+0.02}_{-0.03}$&$-0.54^{+0.09}_{-0.09}$&$+0.071^{+0.003}_{-0.003}$\\
\hline
II&$\log_{10}m_\mathrm{Fe}$&$-3.15^{+0.11}_{-0.12}$&$-3.31^{+0.14}_{-0.16}$&$-3.05^{+0.13}_{-0.11}$&$-3.42^{+0.10}_{-0.10}$&$-2.70^{+0.09}_{-0.09}$&$-2.54^{+0.12}_{-0.13}$&$-2.42^{+0.11}_{-0.12}$&$-2.22^{+0.10}_{-0.11}$&$-2.92^{+0.11}_{-0.12}$&$-3.01^{+0.04}_{-0.04}$\\
II&$[\mathrm{Mg}/\mathrm{Fe}]$&$+0.80^{+0.10}_{-0.09}$&$+0.77^{+0.22}_{-0.20}$&$+0.64^{+0.07}_{-0.06}$&$+0.75^{+0.06}_{-0.05}$&$+0.34^{+0.02}_{-0.02}$&$+0.22^{+0.04}_{-0.03}$&$+0.067^{+0.026}_{-0.006}$&$+0.120^{+0.002}_{-0.002}$&$+0.23^{+0.04}_{-0.03}$&$+0.322^{+0.004}_{-0.004}$\\
II&$[\mathrm{Si}/\mathrm{Fe}]$&$+0.57^{+0.07}_{-0.06}$&$+0.60^{+0.14}_{-0.13}$&$+0.42^{+0.07}_{-0.06}$&$+0.64^{+0.05}_{-0.05}$&$+0.26^{+0.01}_{-0.01}$&$+0.07^{+0.04}_{-0.03}$&$-0.004^{+0.018}_{-0.004}$&$-0.002^{+0.002}_{-0.002}$&$+0.15^{+0.04}_{-0.03}$&$+0.261^{+0.004}_{-0.004}$\\
II&$[\mathrm{Ca}/\mathrm{Fe}]$&$+0.31^{+0.05}_{-0.04}$&$+0.29^{+0.13}_{-0.12}$&$+0.16^{+0.04}_{-0.04}$&$+0.48^{+0.04}_{-0.04}$&$+0.25^{+0.02}_{-0.01}$&$+0.06^{+0.04}_{-0.04}$&$+0.070^{+0.011}_{-0.005}$&$+0.113^{+0.001}_{-0.001}$&$+0.08^{+0.02}_{-0.02}$&$+0.184^{+0.003}_{-0.003}$\\
II&$[\mathrm{Cr}/\mathrm{Fe}]$&$+0.32^{+0.09}_{-0.08}$&$+0.04^{+0.41}_{-0.73}$&$-0.56^{+0.33}_{-0.62}$&$-0.13^{+0.09}_{-0.10}$&$+0.03^{+0.11}_{-0.24}$&$-0.09^{+0.11}_{-0.17}$&$-0.04^{+0.86}_{-0.84}$&$+0.07^{+0.01}_{-0.01}$&$+0.05^{+0.06}_{-0.08}$&$-0.09^{+0.01}_{-0.01}$\\
II&$[\mathrm{Mn}/\mathrm{Fe}]$&$-0.15^{+0.14}_{-0.15}$&$-0.20^{+0.26}_{-0.44}$&$-0.07^{+0.13}_{-0.15}$&$-0.29^{+0.10}_{-0.12}$&$-0.42^{+0.02}_{-0.03}$&$-0.07^{+0.07}_{-0.09}$&$-0.21^{+0.02}_{-0.02}$&$-0.11^{+0.01}_{-0.01}$&$-1.26^{+0.38}_{-0.52}$&$-0.248^{+0.007}_{-0.007}$\\
II&$[\mathrm{Ni}/\mathrm{Fe}]$&$+0.03^{+0.05}_{-0.05}$&$+0.03^{+0.12}_{-0.10}$&$-0.11^{+0.04}_{-0.05}$&$-0.00^{+0.03}_{-0.03}$&$-0.012^{+0.009}_{-0.008}$&$-0.26^{+0.04}_{-0.06}$&$-0.118^{+0.014}_{-0.003}$&$-0.128^{+0.001}_{-0.001}$&$+0.01^{+0.03}_{-0.02}$&$+0.042^{+0.003}_{-0.003}$\\
\hline
$\tau_\mathrm{sfh}$&&$0.46^{+0.12}_{-0.13}$&$2.25^{+1.46}_{-0.80}$&$2.88^{+0.28}_{-0.31}$&$0.97^{+0.17}_{-0.17}$&$0.79^{+0.23}_{-0.14}$&$2.31^{+0.33}_{-0.31}$&$0.27^{+0.57}_{-0.05}$&$0.14^{+0.02}_{-0.02}$&$2.10^{+0.31}_{-0.31}$&$0.82^{+0.16}_{-0.15}$\\
$t_\mathrm{max}$&&$4.68^{+0.91}_{-0.91}$&$10.09^{+1.87}_{-1.95}$&$13.01^{+0.72}_{-1.26}$&$3.77^{+0.79}_{-0.74}$&$5.25^{+0.86}_{-0.95}$&$13.01^{+0.70}_{-1.31}$&$9.50^{+3.17}_{-1.94}$&$3.17^{+0.61}_{-0.50}$&$7.33^{+1.24}_{-1.26}$&$1.83^{+0.33}_{-0.33}$\\
$\tau_\star$&&$26.43^{+14.58}_{-10.37}$&$88.05^{+81.16}_{-52.68}$&$23.70^{+8.47}_{-6.93}$&$13.52^{+5.66}_{-4.11}$&$14.06^{+4.96}_{-3.34}$&$23.19^{+8.78}_{-6.51}$&$4.35^{+15.42}_{-1.54}$&$1.89^{+0.74}_{-0.53}$&$7.56^{+2.46}_{-1.93}$&$0.75^{+0.20}_{-0.14}$\\
$\eta$&&$138.65^{+27.01}_{-24.02}$&$99.88^{+27.59}_{-29.33}$&$47.05^{+11.84}_{-8.83}$&$42.04^{+10.76}_{-8.89}$&$28.82^{+6.92}_{-5.97}$&$32.74^{+9.14}_{-7.32}$&$33.62^{+11.00}_{-8.38}$&$23.48^{+6.39}_{-5.37}$&$4.84^{+1.46}_{-1.29}$&$0.03^{+0.04}_{-0.02}$\\
$\tau_\mathrm{dep}$&&$0.19^{+0.09}_{-0.07}$&$0.81^{+1.37}_{-0.48}$&$0.50^{+0.11}_{-0.11}$&$0.32^{+0.08}_{-0.07}$&$0.47^{+0.15}_{-0.08}$&$0.70^{+0.15}_{-0.12}$&$0.13^{+0.35}_{-0.03}$&$0.08^{+0.01}_{-0.01}$&$1.38^{+0.36}_{-0.25}$&$1.18^{+0.30}_{-0.23}$\\
$t_\mathrm{D}$&&$0.13^{+0.05}_{-0.05}$&$0.17^{+0.05}_{-0.04}$&$0.20^{+0.04}_{-0.04}$&$0.21^{+0.04}_{-0.04}$&$0.07^{+0.08}_{-0.02}$&$0.16^{+0.05}_{-0.05}$&$0.25^{+0.05}_{-0.15}$&$0.26^{+0.04}_{-0.04}$&$0.048^{+0.013}_{-0.006}$&$0.21^{+0.04}_{-0.04}$\\
$f_\mathrm{mix}$&&$0.97^{+0.02}_{-0.03}$&$0.97^{+0.02}_{-0.03}$&$0.990^{+0.007}_{-0.015}$&$0.976^{+0.009}_{-0.012}$&$0.92^{+0.02}_{-0.02}$&$0.989^{+0.008}_{-0.019}$&$0.91^{+0.02}_{-0.02}$&$0.958^{+0.009}_{-0.011}$&$0.985^{+0.006}_{-0.010}$&$0.936^{+0.007}_{-0.006}$\\
\end{tabular}
\label{tab::lg_fullparams}
\end{table*}

\bsp	
\label{lastpage}
\end{document}